\documentclass[sn-mathphys,Numbered]{sn-jnl}% Math and Physical Sciences Reference Style
\usepackage{graphicx}%
\usepackage{multirow}%
\usepackage{amsmath,amssymb,amsfonts}%
\usepackage{amsthm}%
\usepackage{mathrsfs}%
\usepackage[title]{appendix}%
\usepackage{xcolor}%
\usepackage{textcomp}%
\usepackage{manyfoot}%
\usepackage{booktabs}%
\usepackage{algorithm}%
\usepackage{algorithmicx}%
\usepackage{algpseudocode}%
\usepackage{listings}%

\usepackage{hhline}
\usepackage{adjustbox}
\usepackage{makecell}

\usepackage{array}
\usepackage[space]{grffile}
\usepackage{latexsym}
\usepackage{longtable}
\usepackage{tabulary}
\usepackage{booktabs,array,multirow}
\usepackage{amsfonts,amsmath,amssymb}
\usepackage{natbib}
\usepackage{url}
\usepackage{hyperref}
\hypersetup{colorlinks=false,pdfborder={0 0 0}}
\usepackage{etoolbox}
\usepackage[english]{babel}
\usepackage[normalem]{ulem}
\usepackage{algorithm}
\usepackage{xcolor}

\definecolor{codegreen}{rgb}{0,0.6,0}
\definecolor{codegray}{rgb}{0.5,0.5,0.5}
\definecolor{codepurple}{rgb}{0.58,0,0.82}
\definecolor{backcolour}{rgb}{0.95,0.95,0.92}

\lstdefinestyle{mystyle}{
    backgroundcolor=\color{backcolour},   
    commentstyle=\color{codegreen},
    keywordstyle=\color{magenta},
    numberstyle=\tiny\color{codegray},
    stringstyle=\color{codepurple},
    basicstyle=\ttfamily\footnotesize,
    breakatwhitespace=false,         
    breaklines=true,                 
    captionpos=b,                    
    keepspaces=true,                 
    numbers=left,                    
    numbersep=2pt,                  
    showspaces=false,                
    showstringspaces=false,
    showtabs=false,                  
    tabsize=2
}

\newcommand{\ceil}[1]{\left\lceil #1 \right\rceil}

\theoremstyle{thmstyleone}%
\theoremstyle{thmstyletwo}%

\theoremstyle{thmstylethree}%

\begin{document}

\title[Ultra-fast traveltime parameters search by a coevolutionary optimization approach using graphics processing units]{Ultra-fast traveltime parameters search by a coevolutionary optimization approach using graphics processing units}


\author[1]{\fnm{José} \sur{Ribeiro}}\email{joseribeiro1017@gmail.com}
\equalcont{These authors contributed equally to this work.}

\author[1]{\fnm{Nicholas} \sur{Okita}}\email{nicholas.okita@gmail.com}
\equalcont{These authors contributed equally to this work.}

\author[1]{\fnm{Tiago} \sur{A. Coimbra}}\email{tgo.coimbra@gmail.com}
\equalcont{These authors contributed equally to this work.}

\author[1]{\fnm{Jorge} \sur{H. Faccipieri}}\email{jorgehfj@unicamp.br}
\equalcont{These authors contributed equally to this work.}

\affil[1]{\orgdiv{HPG / CEPETRO}, \orgname{UNICAMP}, \orgaddress{\street{Cora Coralina, 350 }, \city{Campinas}, \postcode{13.083-896}, \state{S\~ao Paulo}, \country{Brazil}}}


\abstract{The search for traveltime parameters is a global optimization problem. Several metaheuristics have been proposed to locate the global optima to compute the least amount of their objective functions. However, the theoretical limitations imposed by the no-free-lunch theorem restrict the optimality of such metaheuristics. To escape those limitations, we propose a coevolutionary approach called evolution by neighborhood similarity, which is outside the scope of the restrictions of this theorem and allows us to speed up the search convergence. The technique's effectiveness is based on the approach to exchanging the best individuals found between suitable domains during a differential evolution metaheuristic execution. Moreover, we further expand our technique to graphics processing units, allowing us to explore the performance and memory aspects of the method. Ultimately, our complete coevolutionary algorithm can speed up the parameter search by more than five times and reduce the energy consumption by more than thirty-three times compared to a regular metaheuristic implementation. Although some overheads are still problematic in the technique, we present a first approach that effectively explores the data redundancy hidden in the estimation process, allowing us to provide qualitative, performance, and scalable improvements.}




\maketitle

\section{Introduction}\label{sec1}

Over the years, many traveltime operators were developed for multiple purposes~\citep[see][]{Faccipieri_2018}, from simply improving a seismic-response dataset's signal-to-noise ratio to interpolation and regularization of such datasets to increase their number of traces with robustness and accuracy. Such operators rely on parametric equations based on the seismic response to provide adequate approximations to wavefront kinematic events, such as diffractions and reflections, on a dataset. In that context, the first one was the common midpoint (CMP) traveltime operator used for seismic-response dataset enhancement~\citep{Mayne1962} through rebuilding the dataset in the zero-offset (ZO) gather with enhancement to the signal-to-noise ratio use of coherent information belonging to sets of traces that share the same midpoints. Despite its gains, the amount of data used is quite limited due to its midpoint limitation. More robust methods, based on multi-coverage subsurface dataset analysis such as common-reflection-surface (CRS), nonhyperbolic CRS, and multi-focusing \citep{Jaeger2001,Fomel_2012,Gelchinsky_1999} traveltime operators, mitigate such limitation by using the offset information and information from neighboring traces through midpoint gaps, providing more accurate approximations due to increased redundancy. However, both techniques are also limited to the ZO gather, lacking the generalization present in the finite-offset (FO) gather. This leads to the emergence of two other operators: FO common-reflection-surface (FO-CRS)~\citep{Zhang2001} and offset-continuation-trajectory (OCT) stacking~\citep{Coimbra_2016} traveltime operators, each with its own strengths.

Despite all differences between the traveltime operators, they all rely on parameter estimation to provide accurate approximations of seismic-response events. Such estimated parameters usually have a physical meaning, e.g., wavefront curvature and slowness-phase vector projection. To validate these parameters, we use an objective function by coherence or utility, which in our case, the Semblance~\citep{Neidell_1971} is most common. According to Semblance, the set of parameters that maximize its value for a given dataset point should be the most accurate. It is to be noted that the coherency function relies on data redundancy. Therefore more data is required in more complex traveltime operators to calculate its parameters accurately, thus making the operation more computationally expensive. Furthermore, there may be different sets of parameters for other operators, which also changes how costly it is to compute the function. For example, the CMP method only requires a single velocity parameter on a seismic-response dataset from line acquisition. In contrast, the FO-CRS requires five parameters: two slopes and three wavefront curvatures.

The traveltime-parameter search is mainly a non-convex optimization problem by nature. Therefore, local optimization methods may not be appropriate. Global optimization methods are the most suitable for dealing with the problem more efficiently and avoiding such limitations. In addition, stochastic metaheuristic algorithms have an acceptable trade-off between loss of quality and computational cost to global searches where the cost function is a black box \citep{Spall2003}. In literature, several works applied some metaheuristic based on stochastic optimization to solve the traveltime-parameter estimation; some of them are simulated annealing (SA), very fast simulated annealing (VFSA), differential evolution (DE), particle swarm optimization (PSO), and adaptive differential evolution (JADE)~\citep{Garabito_2001,Cruz_2017,Barros2015,Barros2019,Walda2017,Araujo2019-un,Ribeiro2020}. Qualitatively, all these approaches obtain the traveltime parameters satisfactorily. However, all these methods are limited by the no free lunch (NFL) theorem optimization~\citep{Wolpert_1997,Schumacher_2001}. In other words, they pose a convergence challenge that requires many iterations to achieve desirable results and does not exploit the redundant information present in the estimation process. On the other hand, \citet{Pisnitchenko_2021} explore the redundant information around the measurement point. However, due to a decimate process, that approach may have problems estimating a traveltime that requires many traveltime parameters. Also, the central processing unit (CPU) implementation makes search times prohibitive for pre-stack datasets regularization and enhancement cases.

In order to work around all the mentioned problems, we propose a coevolutionary optimization NFL-free~\citep{Wolpert_2005} called evolution by neighborhood similarity (ENS). A coevolutionary algorithm can tackle the curse of dimensionality using a divide-and-conquer method which separates the search space into subspaces of lower dimensionality. Such an approach decomposes the decision vector into groups of variables that can be optimized cooperatively in cycles. After each cycle, the information gained in the separate optimization steps is joined for the next iteration \citep[][]{Chen_2010}. Based on a coevolutionary strategy, ENS shows higher efficiency in solving a large-scale global optimization problem with a high-performance framework. The main idea of our approach is to find an optimal solution in the interactive domains sequentially divided into parts with the core of some metaheuristics, for which we used the JADE algorithm~\citep{jade}, which is a variant of the DE algorithm~\citep{Storn_1997}. As a result, the dimension and complexity of the problem decrease, allowing ENS to achieve better qualitative results with fewer Semblance computations than a pure JADE execution. Besides, while the approach is platform agnostic, we use a graphics processing unit (GPU) implementation in our numerical experiments to exploit the embarrassingly parallel property inherent in the parameter estimation problem. This difference between the ENS technique and the ENS-GPU implementation is relevant to this paper. The technique focuses on decreasing the number of Semblance computations necessary to achieve suitable qualitative results, while the implementation focus on effectively mapping the technique for the available hardware (either CPU, GPU, or any other), taking advantage of the memory hierarchy and the compute infrastructure (CPU threads or CUDA cores). Because of that, qualitative results are associated with the ENS technique as they are independent, with some reservations regarding the floating point arithmetic, from the hardware used. In contrast, performance results are associated with the implementation as they are directly impacted by the parallel resources available and the overhead imposed by the hardware limitations.

We organize the rest of the work as follows: The traveltime parameter search section describes the general idea behind parameter estimation guided by the adaptive differential evolution metaheuristic and its steps. Also, we argue how the NFL theorem acts on the traveltime-parameter searching problem. The approach by a coevolutionary algorithm section defines how our multi-population coevolutionary algorithm works to avoid the consequences of the NFL theorem, together with some theoretical explanations about the process. The computational aspects section shows how the traveltime parameter search problem is converted into a computational problem, where the multiple steps involved to solve the problem in a distributed fashion and the multiple data dependencies and treatments applied, besides the steps we took to derive our new technique and the main components of ENS. 
The numerical experiments section describes the multiple experiments that effectively compare our ENS technique and implementation against the JADE metaheuristic, such as qualitative, quantitative, performance, and scalability analysis. The critics and limitation section presents the main weaknesses of our implementation, especially involving some convergence and scalability limitations. Finally, the discussion section brings all the previous pieces together and explores the impact of data movement overhead on our executions.

\section{Traveltime parameter search}

In general, the stacking process for interpolation or enhancement of a sample is done with the help of traveltime operators. The traveltime operators are curves or surfaces constructed with the help of parameters that we call traveltime parameters. Usually, producing an objective function using traveltime operators uses specific mathematical operations with the amplitude values of the time sample windows. Such operations generate curves or surfaces that pick amplitude values in samples within the seismic-response dataset. Traveltime parameter search is the process of finding the set of parameters that optimizes an objective function for a referential time sample. The problem of finding parameters for a traveltime operator is of at least fourth-order complexity; in mathematical terms, we have
\begin{equation}
{\cal O} (P_d\times W_l\times T_a\times N_f)\, ,
\label{eq:order}
\end{equation}
where $P_d$ is the total sample points searching in the dataset, $W_l$ is the time-windows length, $T_a$ is the total number of traces in a measurement aperture, and $N_f$ is the necessary number of searches to arrive at an optimum. However, the complexity can be increased according to the objective function utilized. As already mentioned, Semblance has been the most common objective function that makes use of a coherence analysis for such a purpose due to the trade-off between computational cost and quality of results. Also, an efficient implementation of Semblance has $W_l\times T_a$ in order of complexity. Motivated by reducing the $N_f$ factor of eq.~({\ref{eq:order}}), several researchers have sought insight into how best to match optimization algorithms to their problems. The evolutionary algorithms (EA) approach is a powerful tool for solving various optimization problems with various complexity levels. The metaheuristic JADE is an EA, and its application in traveltime parameter search has proved effective \citep{Ribeiro2020}. Nevertheless, the JADE metaheuristic, like any other similar for that specific problem, which searches for one sample at a time, can be limited by the NFL theorem, so there is no way to optimize it without losing the quality of the results. Finally, in the literature, it is known that JADE with archives has the best cost-benefit among the famous DE variants \citep[see,][]{Georgioudakis_2020}.

In the traveltime parameter search problem, individuals represent potential solutions to the search problem. The EA manipulates possible solutions by modifying them with mutation operators and choosing the best among them using selection operators within the same domain. In sample-by-sample searching, a domain is the search space for a time sample. In a domain, when optimizing the Semblance function, an individual is the vector of values of the arguments to the Semblance, i.e., the traveltime parameters. We performed the mutations by applying Gaussian noise vectors instead of the vector combination, and selection may involve choosing those with the highest Semblance value. Since the goal of the search is to find an optimal argument vector, one can typically equate the potential solution with the individual. In our coevolutionary algorithm, individuals serve the same mechanistic purpose as in any EA: they are the fundamental units manipulated by the search operators themselves. Mutation and crossover among domains of interactive domains modify individuals, and selection chooses the best for each domain. Interactive domains are a set of domains where information is exchanged between them. This process encodes the outcomes of interactions between the domains, and depending on the domain, individual entities or the interaction itself may receive values due to interaction with another domain. An algorithm must then decide how to use these outcomes to determine which entities to promote in the next generation and which entities to demote or discard. Any EA can be used to carry out mutations within the interactive domains, with different EA for each domain if desired. In this work, we use the JADE algorithm for all domains and briefly describe its use below; additionally, for details on implementing the JADE algorithm for searching traveltime parameters, see~\citet{Ribeiro2019, Ribeiro2020}.

\subsection{JADE applied to traveltime parameter search}

\begin{figure}[thb!]
\begin{center}
\includegraphics[width=0.70\columnwidth]{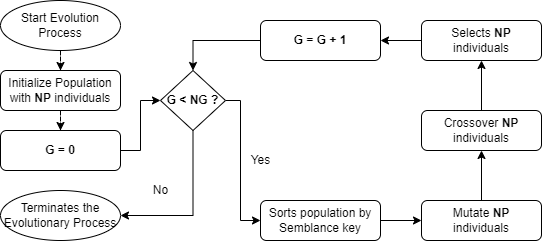}
\caption{{Simplified flowchart of a JADE execution.
{\label{fig:jade_flowchart}}%
}}
\end{center}
\end{figure}

We can summarize the JADE algorithm according to Figure~{\ref{fig:jade_flowchart}}. Naively speaking, in generation zero, i.e., $G=0$, an initial population with $NP$ individuals is generated randomly in the domain and thus transformed over $NG$ generations by mutation, crossover, and selection steps. Finally, at the end of the last generation, the individual that maximizes Semblance is returned as the optimal vector of parameters.

\begin{figure}[thb!]
\begin{center}
\includegraphics[width=0.84\columnwidth]{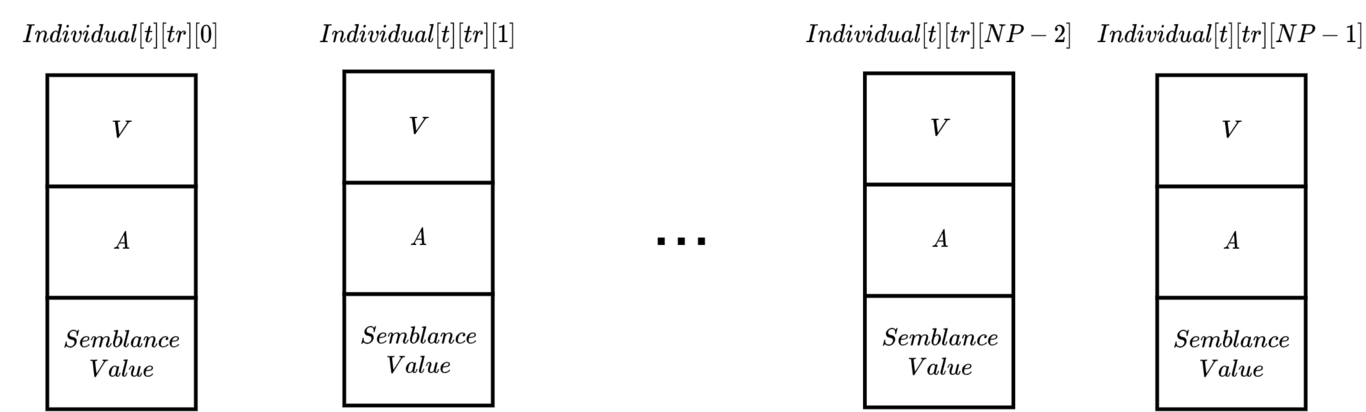}
\caption{{Population Individuals for an OCT traveltime parameter searching.
{\label{fig:population_individuals}}%
}}
\end{center}
\end{figure}

The parameter searching problem aims to find the set that maximizes the Semblance function for a specific index-vector point $(t, tr)$ related to time sample $t=0,...,(NT-1)$ and seismic trace $tr=0,...,(Ntr-1)$. Notably, each trace has a total of $NT$ time samples, which implies a total of $NT$ estimations to accomplish per trace. Therefore, an output dataset with $Ntr$ traces returns us $P_d = NT\times Ntr$ samples searching. Using JADE, each sample estimation is forced to start with $NP$ individuals and then undergo the evolution process. Those individuals are sets of traveltime parameters merged with their respective Semblance values, as exemplified by Figure~\ref{fig:population_individuals}. We use the domains generated by the OCT stacking throughout the explanation to illustrate the search process. In this domain, the individuals have two coordinates, the slope and average velocity presented by the letters A and V, respectively.

\subsubsection{Initializing}

Before the evolution process, the individuals of each population must have their parameters initialized. For example, after providing lower and upper boundary ranges for each traveltime parameter, individuals are randomly generated within the respective domains, and the initialization process is performed. Once the parameter values are specified (for example, A and V), the Semblance function is called to obtain the corresponding value within the generated domain, guiding the population toward convergence during the next steps. To illustrate this process, Figure~{\ref{fig:population_mutation}} shows a population of five initial individuals (four white dots and one yellow dot) initialized under the domain of the corresponding Semblance function.

\subsubsection{Sorting}

Sorting the population by the Semblance value is necessary because JADE uses the $P_{th}$ individual with the most significant Semblance in the population to maximize the problem in fewer iterations. From now onward, we set $P=1$ for simplicity. Figure~{\ref{fig:population_mutation}} shows the best individual depicted in a yellow star.

\subsubsection{Mutation and crossover}

\begin{figure}[thb!]
\begin{center}
\includegraphics[width=0.70\columnwidth]{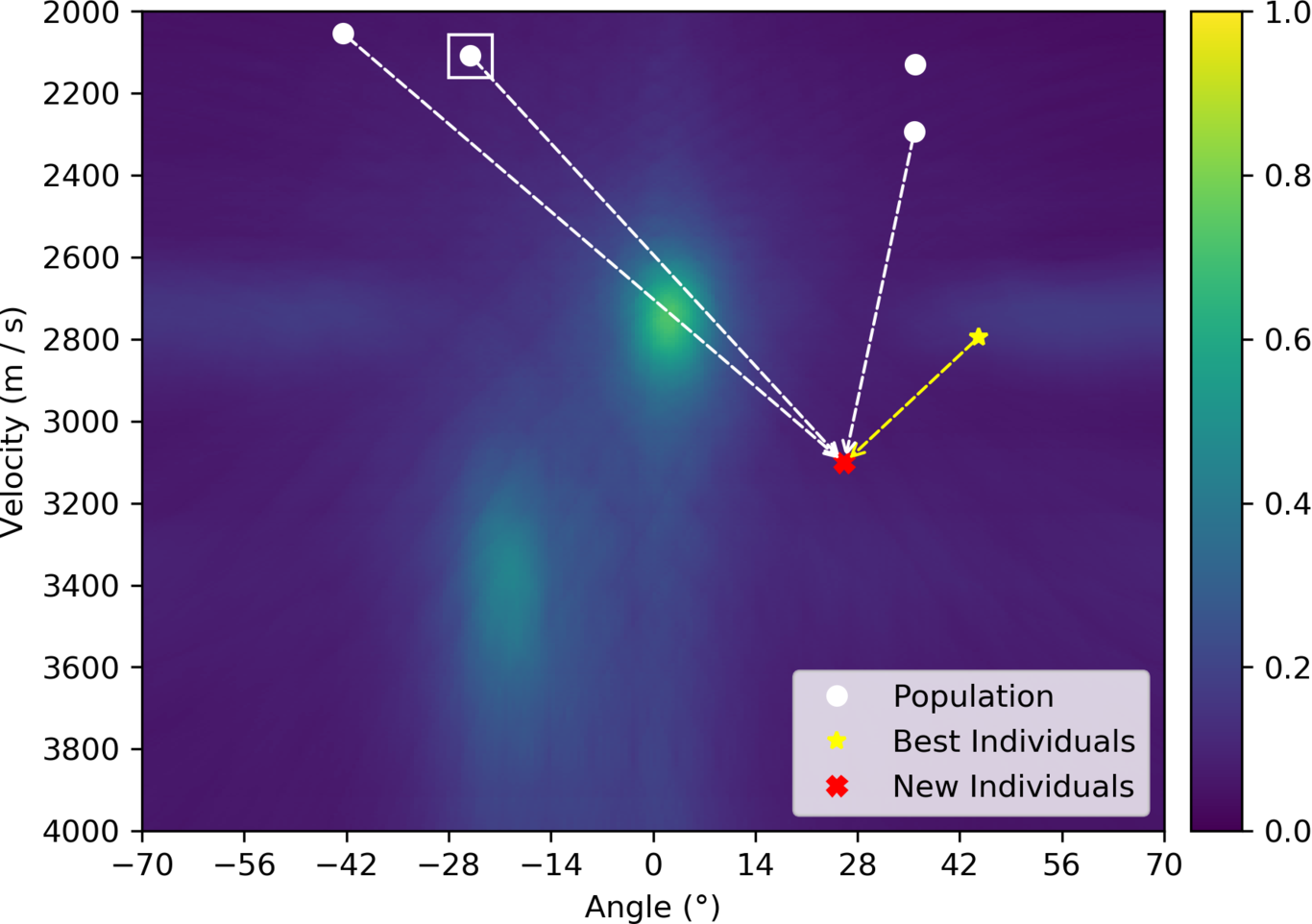}
\caption{Initial Population (white dots), Chosen individual to be mutated (white square), and Child Individual (red cross) generated by the mutation step.}
\label{fig:population_mutation}
\end{center}
\end{figure}

Mutation and crossover generate a new parameter set for each individual in the population. The creation of the new $i_{th}$ individual is simply a linear combination of the $i_{th}$ member, the $P_{th}$ best individual found in the previous sorting step, and two other randomly selected individuals. In Figure~{\ref{fig:population_mutation}}, the $i_{th}$ individual is represented by the squared white dot, the $i_{th}$ child individual by the red cross, the $P_{th}$ best one by the yellow star, and the two random individuals by the non-squared white circles. More technically, the mutant vector is created using one of the DE mutation strategies, namely {\it DE/random-to-pBest/1}. However, our $P=1$ approach makes our JADE implementation a {\it DE/random-to-best/1} mutation strategy \citep[see, ][for more detail]{Brest_2015}. 

\subsubsection{Selection}

Once all previous steps of mutation and crossover are accomplished (Figure~\ref{fig:population_mutation}), a new population (red cross) is generated while keeping the current one (white dots) in memory for the selection step. We cannot continue the execution with both since only one population is expected to evolve during the whole process. To solve that, we compare the $i_{th}$ individual (white squared dot) with its $i_{th}$ child (red cross), choosing the one with the most significant Semblance value and discarding the other. Then the selected individuals are used to compose the next generation's population. Finally, all the control parameters used are the same as \citet{Ribeiro2020}, which uses a historical memory (archived process) to adapt the control parameters \citep{Brest_2006, jade, Leung_2012}. Therefore, unlike a traditional DE implementation, our JADE implementation does not suffer performance degradation due to repeating values in the search. 

\subsubsection{Complete JADE execution}

\begin{figure}[thb!]
\begin{center}
\includegraphics[width=1.00\columnwidth]{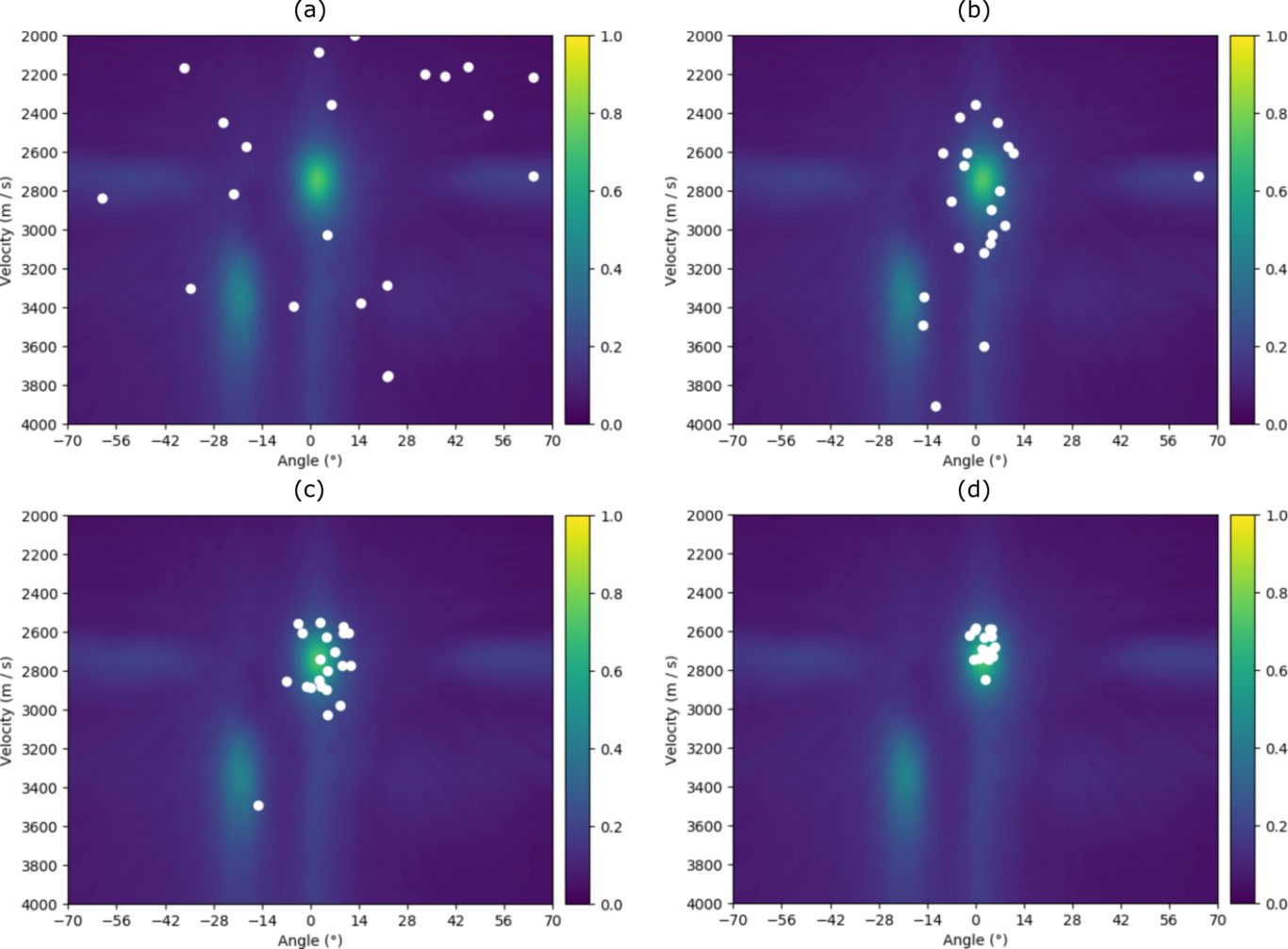}
\caption{{Complete JADE execution with an initial population of twenty individuals (white dots) mutated through (a) Generation zero, (b) Five generations, (c) Ten generations, and (d) Sixteen generations.
{\label{fig:population_jade_execution}}%
}}
\end{center}
\end{figure}

We present a complete execution with a general overview of the JADE steps. First of all, the population of 21 individuals is initialized with random values, as shown in Figure~{\ref{fig:population_jade_execution}}~(a). As illustrated, individuals are well dispersed throughout the objective function space, and no one is considerably near the global maximum. Figure~{\ref{fig:population_jade_execution}}~(b) shows the state after five generations, with individuals starting to concentrate around the function peak. Figure~{\ref{fig:population_jade_execution}} (c) depicts five more generations in which acceptable parameters are obtained. Finally, Figure~{\ref{fig:population_jade_execution}}~(d) shows the execution after sixteen generations, where the best solution is reached. Though the satisfactory result, repeated executions may imply different outcomes, mainly due to the random nature of the metaheuristic. Besides, the efficiency of a search process in all population-based nature-inspired algorithms depends on exploration and exploitation. Figures~\ref{fig:population_jade_execution}~(a)-(b) show the exploration process of new undiscovered regions of the search space in each generation. In contrast, Figures~\ref{fig:population_jade_execution}~(c)-(d) show the exploitation process directing the search toward known reasonable solutions. Based on \citet{Vo2001}, we say that the global convergence of most metaheuristics has not yet been fully proven from a theoretical point of view.
While there are even convergence results regarding the quality of the solution for some classes of metaheuristics, when appropriate probabilistic assumptions are made, they are not very useful in practice, as disproportionate computation time is often required to achieve these results. Generally, the convergence guarantee is achieved for computation time tending to infinity. Therefore, EA or any similar algorithm has no guarantee of convergence toward the global optimum with a number of iterations less than a certain threshold. 

\subsubsection{JADE Semblance computations}

Considering that the Semblance calculation is the core of the EA applied to the traveltime parameter searching problem, we calculate how often that operation is called during execution. The number of Semblance function calls for all domains in a specific dataset is calculated as
\begin{equation}
\label{eq:jade_semblance}
num\_jade\_semblances\_calls\ =P_d\times NP\times (1 + NG)\, .
\end{equation}

\subsection{Restrictions imposed by NFL to the traveltime parameter search}

As mentioned, the traveltime parameter search is a black box arbitrary search problem in a real-world situation.
Although there is practical progress in using global stochastic search techniques, such methods can have theoretical limitations by the NFL implications. 
Therefore, there is no scope for distinguishing the performance of these biased sampling methods for all plausible seismic-response dataset types. The fundamental limitation of these search algorithms is that there is no a priori knowledge of the interest domain problem. Therefore, the algorithm cannot exceed the best search performance if no domain-specific knowledge is used to select an appropriate cost-function-set representation.

In order to begin our analysis, we use the following expression to summarize the NFL theorem statement: No search algorithm is better than another for all possible performance measures when its performance is averaged over all possible discrete functions. Here, given a fixed number of points in the search space, the performance measure adopted is the one that measures the number of steps that are necessary to find a solution to a given cost-value. However, before formulating the NFL theorem questions about traveltime parameter search, it is essential to introduce some terminology to reduce the scope for misinterpretation. Given a traveltime operator, we characterize its traveltime parameter search problem by a search space, ${\cal P}$, the set of objects over which the search is conducted, and an absolute-normalized objective function, $f$, which is a mapping from ${\cal P}$ to the space of the objective function in the interval value $[0,1]$, i.e.,
\begin{equation}
f:{\cal P} \rightarrow [0,1]\, .  
\label{eq:f}
\end{equation}
Pay attention that all the traveltime parameters that form the set ${\cal P}$ are possible values with physical meaning to compute the traveltime operator under analysis.

In order to make our analysis more transparent, let's assume, for simplicity, that the objective function adopted here is the Semblance function, which is represented as
\begin{equation}
S:{\cal P}\times{\cal D}\times{\cal W} \rightarrow [0,1]\, ,   
\label{eq:S}
\end{equation}
where ${\cal W}$ is the set of all possible Semblance adjustment parameters (e.g., time window, coordinate-displacement aperture, traveltime initial coordinates, etc.), and ${\cal D}$ is the set of all possible seismic response datasets. Therefore, from eq.~(\ref{eq:f}) and~(\ref{eq:S}), we have
\begin{equation}
f(p) = S(p;\text{\#data},\text{\#config})\, , 
\label{eq:Sf}
\end{equation}
where $p\in{\cal P}$, \#data $\in{\cal D}$, and \#config $\in{\cal W}$. Besides, we restrict attention to decimation (downsampling) in which a physically acceptable discretization in search space ${\cal P}$, though perhaps quite large, is finite. To the co-domain, the restrictions are automatically met for digital computers, typically with some 32- or 64-bit representation of the zero to one range of real numbers.

Although the set of all Semblances functions is very large, it may happen that it is not large enough to contain all functions of type in eq.~(\ref{eq:f}). However, there is no need to contain all possible functions but a subset with specific properties. For that, \citet{Schumacher_2001} present the sharpened NFL theorem, which can summarize as follows: The NFL theorem holds for a set of functions if and only if that set of functions is closed under permutation. In other words, no structure can be used for search if a class of functions does not change by any permutation on the input space. Hence, all search strategies show the same behavior. The argument in our notation says for each pair (\#data$_*$,\#config$_*$) in ${\cal D}\times{\cal W}$. It must be possible to find a finite subset \{(\#data,\#config)\}$_N$ contained in the set \{(\#data,\#config)\}$_\lambda$ to which all permutations of the fitness function $f_*(p) = S(p;\text{\#data}_*,\text{\#config}_*)$ belong (for the assumed ${\cal P}$ and $[0,1]$ discretization). Therefore, there is no best algorithm for searching in such a set in the affirmative case.

On the other hand, in the case some subset is not closed under permutation, \citet{Droste_2002} making use of the almost NFL (ANFL) theorem, show that for any function for which a given algorithm is effective, related functions exist for which the performance measures of the same algorithm are substantially worse. In other words, the ANFL theorem implies that a search strategy has to pay for its success in one class of functions with its bad behavior in many other classes of functions that may have different or lesser complexity than the one. Therefore, each search strategy has some intuition about the optimization functions; however, such intuitions are penalties for other functions. For example, an algorithm that excels at low noise conflicting dips may have problems with mono events and high non-coherent noise. 

Finally, analyzing NFL theorems forces us to consider which problems we must solve. However, without any information from this set, we must be careful in using algorithms for specific class problems, as such algorithms may be bad for other classes of genuine interest. Besides, as a disclaimer, this section aims to clarify the possible difficulties of an EA in the traveltime parameter search using situations where the NFL theorem holds. Therefore it is outside the scope of this work to provide mathematical proof that any traveltime operator under some measure of coherence is subject to the requirements of some version of the NFL theorem.
 
\section{Approach by a coevolutionary algorithm}

The NFL theorem is a fundamental result in the black-box function optimization field. Its most basic and informal form states that all global-search algorithms perform equally well when averaged over a set of functions closed under permutations to be optimized. As already seen, the NFL theorem can be applied to the class of iterative, data-driven search algorithms that optimize an objective function, such as the traveltime parameters searching by sample-by-sample. Nevertheless, the interactive nature of fitness evaluation in co-optimization implies that coevolutionary algorithms are not generally members of this class \citep{Wolpert_2005}. Based on these premises, we developed  a multi-population coevolutionary algorithm in which individuals of one population interact with individuals of other populations. Therefore, in this decomposition strategy, the quality of a solution to the problem involves an interaction among many domains.

Furthermore, given an objective function $f$, the fitness relationship between any two individuals, $p_1$ and $p_2$ in ${\cal P}$, in an EA, is given by comparison between $f(p_1)$ and $f(p_2)$ to measure the best fitness. By contrast, coevolutionary algorithms do not use such a direct comparison of the fitness of functions. Instead, two individuals are compared based on their outcomes from interactions with other individuals. For this reason, the fitness in a coevolutionary algorithm has been called subjective fitness, subject to the changing populations \citep{Popovici_2012}.

Based on the formal notion of an interactive domain by \citet{Popovici_2012}, we define our interactive domain as consisting of fitness functions of the form
\begin{equation}
g:{\cal P}\times\cdots\times{\cal P}\rightarrow [0,1]\, ,    
\end{equation}
where the tuple $(p_1,\cdots,p_n)$ in ${\cal P}\times\cdots\times{\cal P}$ is an interaction vector, the value $g(p_1,\cdots,p_n)$ is an outcome of the interaction, the ordered set $[0,1]$ is the outcome set, and each element $p_i$, $i=1,\cdots,n$, belonging to the domain ${\cal P}$ is an entity. Also, an interactive domain defines many domain roles and entities that play each role. Thus, each objective function, $f_i$ (as in eq.~(\ref{eq:f})), is interpreted as giving an outcome to the entity $p_i$. The optimization is organized in cycles. Each cycle activates the interactive domain in a divide-and-conquer approach over the different domains. The interaction vector is updated using the current best entity of each domain. Therefore, once an interactive domain is defined, a co-search and co-optimization process which assigns values to interactions among entities can be described.

In order to extract the best-fit traveltime parameters from all domains in a seismic-response dataset, we build a set of functions formed by separable mapping. Such separability means that the influence of a variable (entity) on the fitness value is independent of any other variables \citep{Chen_2010,Trunfio_2015}, i.e., maximizing the objective functions $f_i$ is equivalent to maximizing the outcome of the interaction function $g$ and vice versa.
Therefore, we can define the set of $g$-type functions by an average of the functions $f_i$, that is
\begin{equation}
g(p_1,\cdots,p_n) = \frac{1}{n}\sum_{i=1}^{n}f_i(p_i)\, ,
\label{eq:g}
\end{equation}
where $i = 1 + t + NT\times tr$, $n=P_d$, and $f_i$ is a Semblance function over the domain of sample in $(t,tr)_i$.

In brief, the coevolutionary algorithm idea consists of decomposing the original high-dimensional problem (interaction domain) into a set of lower-dimensional subproblems (domains) which are easier to solve. Typically, each subproblem is assigned a subpopulation of candidate solutions (entities), which is evolved according to the adopted optimization metaheuristic. During the process, only cooperation happens in the fitness evaluation through an exchange of information between subpopulations \citep{Trunfio_2015}. Therefore, the main idea of our algorithm is to explore the similarity between the domains that are close together. Then the limitations of the NFL theorem are circumvented through a process of information exchange between domains.

\subsection{The NFL-free algorithm}

The main idea about our coevolutionary algorithm is to find a kind of self-play structure \citep{Wolpert_2005} to make it superior to others. In other words, we optimize the functions $f_i$ by transforming this problem into a coevolutionary approach in order to design a set of $g$-type fitness functions where the NFL theorem does not hold. From that, we design a procedure that exploits the obtained free lunches. Furthermore, given a function set where the NFL theorems don't hold, we say an algorithm is NFL-free when exploiting some free lunch from such a set.

To explain our reasoning, suppose a set of functions of type $f_i:\{0,1\} \rightarrow \{0,1\}$. The most straightforward set closed under permutation, based on that type function, is 
\begin{equation}
{\cal F} = \left\{\begin{array}{ccc}
f_1 &=& \langle 0, 1 \rangle; \\
f_2 &=& \langle 1, 0 \rangle.
\end{array}\right\}\, .
\label{eq:cF}
\end{equation}
Therefore, the NFL theorem holds in ${\cal F}$, in which the minimal searching required is one. Now, based on eq.~(\ref{eq:g}), we create a function set of the form $g_{ij}:\{0,1\}^2\rightarrow \{0,1/2,1\}$, through the functions in ${\cal F}$ with $n=2$, as
\begin{equation}
{\cal G} = \left\{\begin{array}{ccc}
g_{11} &=& \langle 0, 1/2, 1/2, 1 \rangle; \\ 
g_{12} &=& \langle 1/2, 0, 1, 1/2 \rangle; \\
g_{21} &=& \langle 1/2, 1, 0, 1/2 \rangle; \\ 
g_{22} &=& \langle 1, 1/2, 1/2, 0 \rangle.
\end{array}\right\}\, .
\end{equation}
As the set ${\cal G}$ is not closed under permutation, the NFL doesn't hold. Besides, the best algorithm in this set takes a maximum of two searches. As much as set ${\cal G}$ is non-NFL, the problem is that the number of reliable searches to find an optimum is the same number of searches to find the optimum in two samples of set ${\cal F}$. Therefore, the construction by eq.~(\ref{eq:g}) maintained the complexity of the problem.

In order to reduce the problem's complexity, we assume a similarity hypothesis, i.e., given a dataset with $P_d>1$, there exists some $i \neq j$ such that $f_i = f_j$. Thus, in our simple example, we have a subset ${\cal G}' = \{g_{11},g_{22}\}$, in which just one search is necessary to find an optimal. It is not difficult to see that the idea holds even if we make set ${\cal F}$ more complex. Pay attention to the similarity hypothesis as a dataset condition and is not of set ${\cal F}$, i.e., from eq.~(\ref{eq:Sf}), given two different traveltime initial coordinates inputs, \#config$_1$ and \#config$_2$, implies in $S(p;\text{\#data},\text{\#config}_1)$ equal to $S(p;\text{\#data},\text{\#config}_2)$, which are the same functions in a set ${\cal F}$ but in different $(t,tr)$-sample in a dataset. Furthermore, other structures can be found in a dataset in order to build another set ${\cal G}'$ with more free lunches. However, it is not explored in this work. Finally, as shown below, we empirically justify the similarity hypothesis for the traveltime parameters search case.

\subsection{Similarities between neighboring domains}

Let us empirically explain our algorithm's idea with the following example's help. Figure~{\ref{fig:ens_forward_propagation_example}} shows a grid with domains generated by the Semblance function applied to a synthetic dataset containing around seven hundred time samples and one hundred traces. The visual similarity among those domains is quite evident. For example, when comparing the domain sample $(t=651,tr=22)$ with the one above (the same trace, but the time sample above, i.e., the sample in $(650,22)$), while different, the point below in time also shares remarkable similarities with the main domain, particularly concerning the global maximum position. The same happens after setting the time sample at 651 and comparing the similarities between trace positions 23 and 24. Physically speaking, it is natural that there are no sudden changes between close samples since they share most of the same dataset. Therefore, it is clear that searching parameters using information exchange among these domains undoubtedly helps to improve the results and the speed of convergence of estimates from consecutive traces and time samples, i.e., it reduces the value of the $N_f$ factor of eq.~(\ref{eq:order}). 

Furthermore, we define a trust region as the set of points in the domain where something can be learned for global convergence (e.g., a hill-climbing region). Based on that, we say the similarity between the two domains is when their optimal point belong to each other's trust region. 

\subsection{Information propagation}

\begin{figure}[thb!]
\begin{center}
\includegraphics[width=1.00\columnwidth]{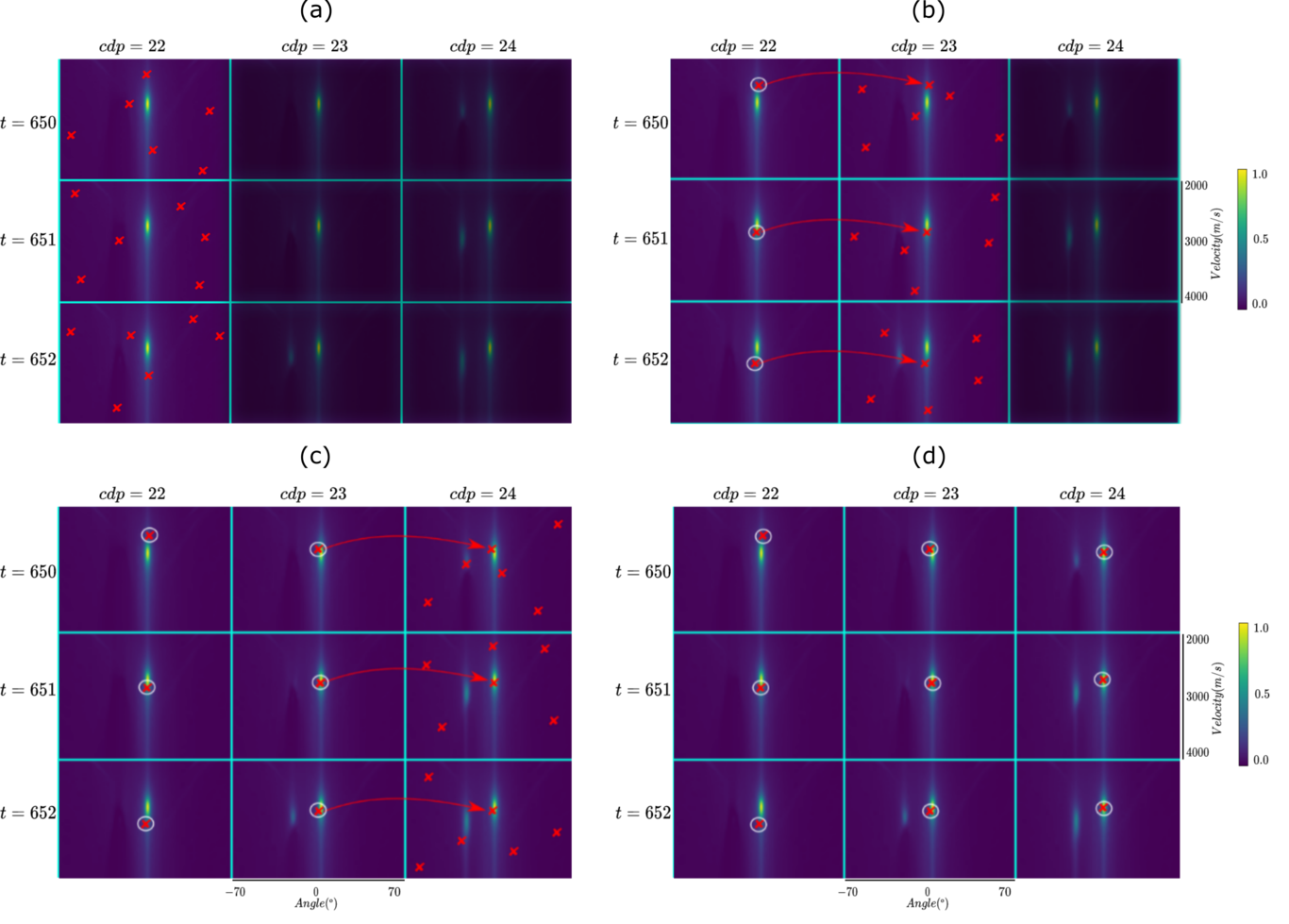}
\caption{{Forward propagation step showing the individual exchange between domains for the (a) First trace, (b) Second trace, (c) Third trace, and (d) Final third trace state.
{\label{fig:ens_forward_propagation_example}}%
}}
\end{center}
\end{figure}

We use an interactive domain scheme to take advantage of the similarities between the domains. The information exchange between domains occurs through forward propagation (from left to right) and backward propagation (from right to left). Figure~{\ref{fig:ens_forward_propagation_example}} illustrates the method in a set of three traces. Figure~\ref{fig:ens_forward_propagation_example}~(a) shows that we start the search with the leftmost trace in this interactive domain. First, the population of $NP$ individuals is initialized for each time sample of the first trace (In our example, $NP = 6$, also, $NG=1$). Then, in the last generation, the individual with the most outstanding Semblance value in each time sample is chosen (Figure~\ref{fig:ens_forward_propagation_example}~(b)). The search for the first trace is completed, and the search for the second trace takes place. However, unlike the first process, we used the best individuals from this first estimate to compose the initial population of this second one, as shown in Figure~\ref{fig:ens_forward_propagation_example}~(b). We propagated the individuals from a previous search to increase the chances of finding the global maximum by adding bias to the evolutionary process. The new individual replaces one of the randomly generated population members, thus keeping the population size at six. After finishing this second JADE generation process, we propagated the best individuals from the second trace to the third trace (Figure~\ref{fig:ens_forward_propagation_example}~(c)). Finally, Figure~\ref{fig:ens_forward_propagation_example}~(d) shows the forward propagation approach's parameter estimation conclusion. Therefore, if the similarity assumption holds for all three domains for the same value of $t$ (as seen in our example), then for the last domain on the right, the search is similar to a JADE with $(NG+1)\times 3$ generations. However, the exploration-exploitation trade-off tends more toward exploration due to restarting the $(NP-1)$ individuals with each change of domain. 

\begin{figure}[thb!]
\begin{center}
\includegraphics[width=1.00\columnwidth]{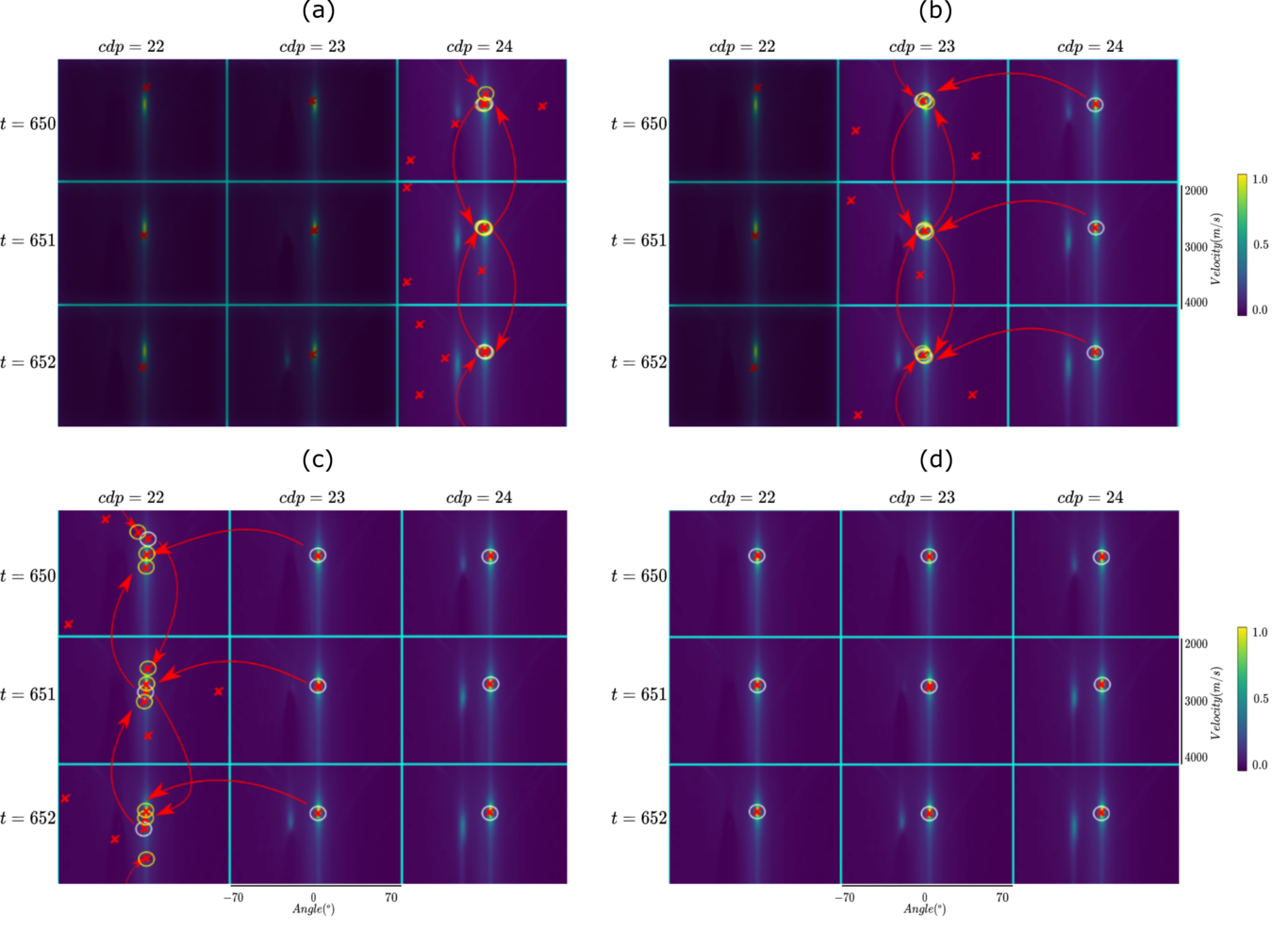}
\caption{{Backward propagation step showing the individual exchange between domains for the (a) Third trace, (b) Second trace, (c) First trace, and (d) Final first trace state.
{\label{fig:ens_backward_propagation_example}}%
}}
\end{center}
\end{figure}

We introduce a backward propagation step to solve the problem created by the forward propagation, in which the algorithm has not yet reached an optimum over the leftmost domains (again, $NP=6$ and $NG=1$). This purpose is to give the leftmost traces, the first estimated ones, the chance to improve their parameter set using information from all previous computations. After applying forward propagation, a new right-to-left estimation occurs. Figure {\ref{fig:ens_backward_propagation_example}} illustrates this new process. Firstly, the rightmost trace has its parameter estimated, as shown in Figure~\ref{fig:ens_backward_propagation_example}~(a). However, due to the forward step, all the time samples of that trace already have the best set of parameters found, which are used to compose the initial population of the backward propagation estimation. Therefore, the best individual found on time sample 651 can be introduced to the 650 and 652 initial populations so that their estimations can be further improved. Conversely, the best individuals from the 650, 651, and 652 samples can be introduced to the initial population of time sample 651. Three individuals are now used to bias the estimation, contrasting with only one from the forward approach. In contrast, three randomly generated individuals must be removed from the population to keep its size fixed at six, creating a population with three random individuals and the other three biased ones obtained from previous estimations. However, this is only valid for the rightmost trace. At least four individuals are available when the other traces are estimated because the previous trace can also contribute to the estimation, similar to the forward scheme. While other neighbor individuals could be used to bias the estimation (for instance, the diagonal neighbors), we maintain only the four biases mentioned above to allow the usage of small populations and keep population diversity. Figures~\ref{fig:ens_backward_propagation_example}~(b) and~(c) illustrate this process for the two last traces. Then, Figure~\ref{fig:ens_backward_propagation_example}~(d) shows the final state of the found individuals after the backward propagation is accomplished. Due to this approach, it is possible to refine the previously estimated parameter sets on top of the forward propagation, thus increasing the chances of finding the global maximum. Therefore, in that process, the exploration-exploitation trade-off tends more toward exploitation due to initialization with the best four individuals. Finally, one more forward step may be necessary to exploit left-to-right traces better. This forward step has a backward behavior, but with the direction reversed. Thus, our algorithm can be summarized in the following steps: After initializing with one forward (more exploration) propagation process, each cycle is characterized by a backward-and-forward (more exploitation) process.   

\subsection{How does our self-play approach work?}

Given a traveltime with a discretized parameter-domain ${\cal P}$ with $2N$ points and the response of the Semblance function (with all the sufficient settings for that) over this domain is a needle-in-a-haystack function, i.e., a function that has only one value greater than zero, everything else is zero. In our example, the objective is to find a search strategy that finds the maximum value in this domain (called Haystack) with the lowest number of searches. However, looking at only that domain, no strategy could beat the random search with archives here (which it needs on average $N$ searches to convergence) because nothing can be learned from one point over another outside that non-zero cost-value since the trust region is just that point. Besides, we can contaminate with uniformly distributed noise with a high signal-to-noise ratio that would not change the nature of the problem.

Let us show how our coevolutionary approach works in a simple case. Suppose an output dataset with $P_d = k$, and the similarity hypothesis holds in these $k\geq 2$ domains (distributed in just one line, i.e., the variable $t$ is constant). Also, let's assume they are all Haystack domains (they may have different noises). In a self-play scenario, where everyone uses the same strategy to decide their actions, the domains interact and change each other's learning environments. Based on that, we expect to see that the cost-value average of each domain should increase at the same rate in the case that the similarity hypothesis holds. Therefore, we use a random search algorithm with archives in each domain to facilitate understanding in this context. In the forward-exploration process, each algorithm performs $N/k_*+1$ searches (by the law of total probability, we can take $1.5 < k_* < k$), i.e., it repeats the best point of the previous domain and performs $N/k_*$ new searches. At the end of that procedure, the process would have performed $N(k/k_*)+1$ different searches and $k-1$ repetitions (there is nothing to copy in the first domain). Since the algorithm has archives, these $N(k/k_*)+k$ searches would have reached the optimum on average. In the backward-and-forward exploitation process, we do one search for each direction as there is nothing to exploit, and with that, we have a total of $2k$ searches in this part. Hence, $N(k/k_*)+3k$ searches are performed within the interactive domain with $k$ domains. Therefore, a typical sample-by-sample search takes $N$ searches to achieve an average optimum for each domain. While our approach spends $N/k_* + 3$ searches for domains. 

Finally, the same behavior occurs in more complex functions as long as the similarity hypothesis holds and the exploration-exploitation relationship has the appropriate search distribution with one or more cycles. To justify that affirmation, suppose a function with some particular property for which the best algorithm makes only $N' < 2N$ searches, on average, to find a global optimal over its trust region of length $2\beta > N'$. Besides, suppose more $(k-1)$-domains with the same trust region but its optimal points in different places over that region. Therefore, $(N-\beta+N')$ searches, on average, are necessary to find an optimal in that function class. If we apply a resampling reduction of the points of $P$ to a $P'$ with $2(N-\beta)+1$ points by a downsampling process that keeps the $2(N-\beta)$ points and transform the $2\beta$ points in one with the maximal block value. We have the same or similar behavior as the Haystack functions in a non-regular grid. The issue here is that it would take, on average, $(N-\beta)$ points to find an optimum in this configuration. Hence, on average, a total number of $(N-\beta)(k/k_*) + k$ searches are performed in the exploration process to find an optimum for each $P'$-domain in the interactive domain. For the exploitation process, once inside the trust region in each $P$-domain and, by assumption, an optimal search algorithm spends $N'$ searches therein. From that, we can assume $(N'+2)k$ searches are sufficient to find the optimum within a trust region. Thus, $(N-\beta)(k/k_*)+(N'+3)k$ searches are realized in the interactive domain with at least $(N - \beta)/k_* + (N'+3)$ searches by domain, which is minor than $N-\beta+N'$. Wherefore, we define as an ultra-fast traveltime parameter search the fact that the new set of extended operations ($g$-type fitness functions) can inherit any free lunch in the coherence functions set. Therefore, no algorithm that only looks at information from a single domain can be superior to our approach. Even if we obtain an NFL-free algorithm on a single-domain, our multi-domain approach can also take advantage of the same free lunches. Hence, changing the paradigm of searching on a single domain to multiple domains creates more robust possibilities for performance improvement.

\subsection{The performance-quality trade-off by an interpolation process}

We introduce an interpolation process between time samples to reduce further the number of searches performed in the proposed method. The idea is to make a linear interpolation between one or two skipped samples and compute the individuals of these domains by the values obtained.

\begin{figure}[thb!]
\begin{center}
\includegraphics[width=1.00\columnwidth]{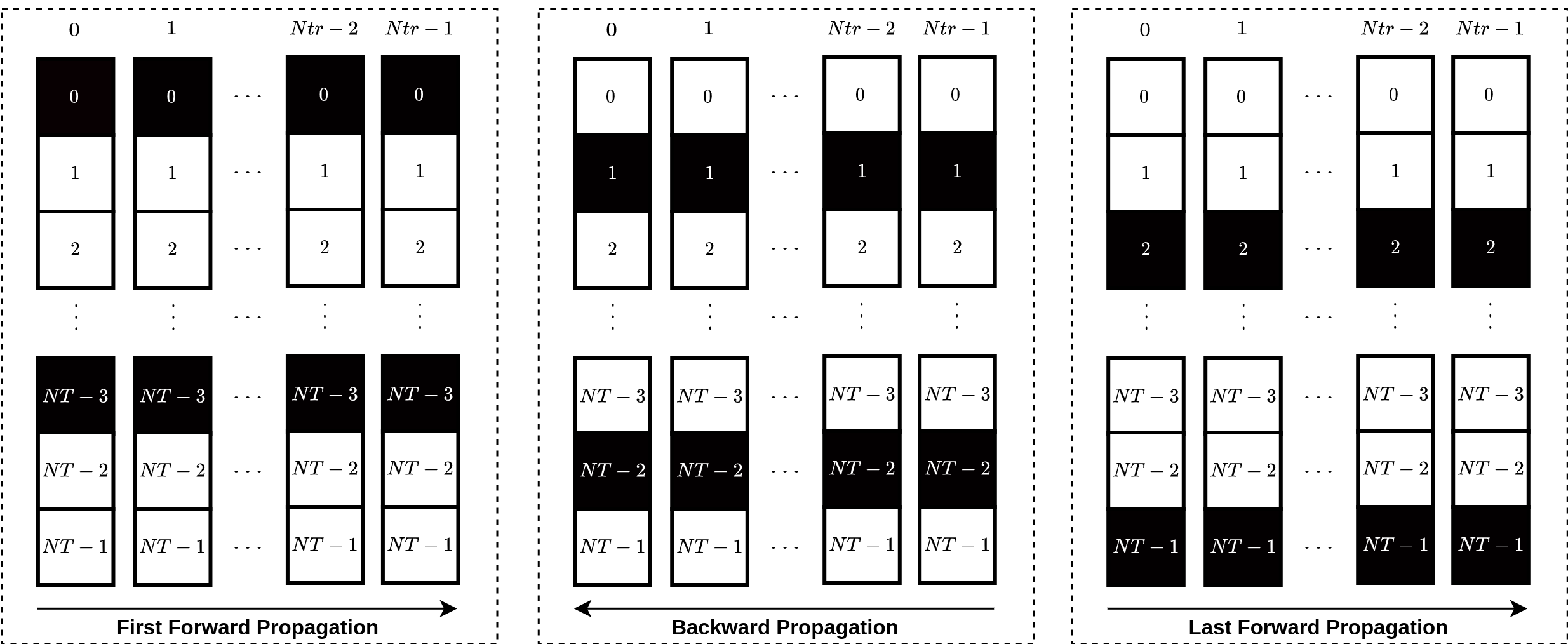}
\caption{{Parameter searching by two-sample skipping showing the estimated samples (black) during each one of the three steps: first forward (left), backward (center), and last forward (right) propagation.
{\label{fig:ens_flowchart}}%
}}
\end{center}
\end{figure}

In summary, there are three situations of interpolation to the method. First, no interpolation. Nothing that has been said so far is changed. Second, the algorithm skips one sample. The forward-exploration process typically occurs in non-skipped samples. At the same time, the skipped samples are interpolated and have their Semblance computed. The backward-and-forward process occurs on the skipped domains, and the interpolation occurs on the remaining ones. Lastly, the algorithm skips two samples. In non-skipped samples, same as in the previous case, the forward-exploration process occurs, while the skipped samples are interpolated and have their Semblance calculated. The backward-exploitation process is performed for the first skipped sample, and the interpolation occurs in the other samples. The forward-exploitation process is performed for skipped second samples, and the interpolation occurs again for other samples, as seen in Figure~\ref{fig:ens_flowchart}.

On the seismic trace, the events have a temporal bandwidth $b\Delta t$ (let's suppose $b = 1,2,\cdots$ samples), and the Semblance coverage has the bandwidth $(b+\gamma)\Delta t$, $\gamma > 1$, which covers the event. Based on this information, we defined the Semblance edge as the temporal sample of the top and bottom of this coverage in the trace. However, by interpolating the samples in time, we have a phenomenon at the Semblance edge over an event. Besides, when traces are similar in that region, i.e., there are no dip events. Such a phenomenon does not occur. However, in dips events, a kind of spatial aliasing can occur, i.e., the entities at the Semblance edge in time direction are suboptimal. Generally, this part of the Semblance coverage does not have vital information. Besides, for each skipped sample, only one Semblance is computed per process through the interpolation of $p_i$ values. So the problem occurs when a sample is being interpolated with one end being noise, and with the following trace shifted, one or two samples with the same interpolation situation. Therefore, the domain in this situation only receives noise and non-ideal sub-optimal values as input in the backward-and-forward process. Finally, if no skip is performed, there is no interpolation process, then no quality loss occurs. Therefore, a performance-quality trade-off is a relationship between the reduction in the total number of searches and the sub-optimality of investigations with the loss of optimal at the Semblance edges. Furthermore, with more cycles or no interpolation, the quality is restored, but the performance is reduced.

\subsection{Evolution by neighborhood similarity}

Given all the previous descriptions, we finally reach the ENS algorithm. The algorithm steps are described in a procedural mode as the following, where the input parameters for the algorithm are: $NP_f$ and $NG_f$ are the population and generation size, respectively, for each domain to the forward-exploration propagation, while $NP_{bf}$ and $NG_{bf}$ are the population and generation size, respectively, for each domain in each step to the backward-and-forward propagation. Also, $N_{skip}$ is the number of time samples skipped for interpolation, and $N_{cycle}$ is the total cycles (backward-and-forward propagation) to be performed. Finally, as mentioned, $NT$ and $Ntr$ are the total time samples and traces number.

Furthermore, in order to avoid temporal aliasing, we recommend $N_{skip} \le 2$. Also, by the performance-quality trade-off, more samples skipped can deteriorate the result, as in dip events. Figure~\ref{fig:ens_flowchart} shows a roadmap of the algorithm using $N_{skip} = 2$ and $N_{cycle} = 1$. In the following paragraphs, we explain the algorithm in more detail.

The algorithm creates an output dataset ordered by offset-midpoint with a total of $P_d = Ntr \times NT$ samples  to keep the similar domains as close as possible. Then, the step is to make the stages list, representing the forward-and-backward information propagation cycles as previously described. In other words, after the first forward-exploration propagation, in sequence, the backward-and-forward propagation is performed according to the $N_{cycle}$ parameter.

The algorithm can be divided into three stages.
In the first stage, the forward-exploration information propagation is done through each of the traces going from the leftmost trace to the rightmost one. On the first trace, the domains $(t_0,0)$ are initialized with $NP_{f}$ individuals each, and within it, some EA is applied (JADE in our implementation) with $NG_{f}$ generations through each of these domains individually, as shown in Figure~\ref{fig:ens_forward_propagation_example}.
Note that $t_0 = 0, N_{skip}+1,\ldots,t = (N_{skip}+1)\times i_t \le NT-1$, and $i_t=0,1,\ldots,\ceil{\frac{NT}{N_{skip}+1}}$, where the operator $\ceil{\bullet}$ represents the ceiling function, which rounds the fractional number to the least integer with a value greater or equal to $\bullet$.
Besides, after processing all domains $(t_0,0)$, an interpolation step generates a temporary result for all domains $(t, 0)$, which were not generated during the mutation process. The interpolation uses the best individuals from the final generation found in domains $(t_0, 0)$. No interpolation is performed if $N_{skip} = 0$. In the following trace, but still, during the same stage, the domains $(t_0, 1)$ initialize its $NP_{f}$ individuals, where at least one of them is the best individual from the final generation of $(t_0, 0)$. The generations are performed in these domains in the same pattern as the previous trace. And again, the domains that were skipped obtain their results by interpolating the final result from the other domains. The process continues until the domains $(t_0, Ntr-1)$ are processed, as shown in the information propagation section. If the user-defined $N_{cycle} \ge 1$, we continue to perform the backward-and-forward propagation stage.

In the second stage, the biggest difference is that the iteration through the traces is in reverse order. So we initialize the populations of the $(t_1, Ntr-1)$ domains, where $t_1 = init,init+(N_{skip}+1),\ldots,t = (N_{skip}+1)\times i_t+init \le NT-1$, and $i_t=0,1,\ldots,\ceil{\frac{NT}{N_{skip}+1}}$; in which $init = 0$ if $N_{skip} = 0$, or $init = 1$ otherwise. The population size generated for this stage is defined by $NP_{bf}$ and the number of generations by $NG_{bf}$. Also, note that the first three individuals are chosen as the best individuals from neighbor domains, and the fourth is chosen as the best in the current one, as shown in Figure~\ref{fig:ens_backward_propagation_example}.
Besides, in case $N_{skip}>0$, an interpolating process occurs for the $(t, Ntr-1)$ samples that were not processed, using information from $(t_1, Ntr-1)$. Since we already have results obtained on the previous step for the interpolated domains $(t_0, Ntr-1)$, we keep the one with a higher coherence value. The following domains $(t_1, Ntr - 2)$ until $(t_1, 0)$ are processed to finish the backward-exploitation propagation step.

In the last stage, which is the second part of the backward-and-forward propagation stage, we iterate again through the subset of traces from the leftmost one to the rightmost.
Hence, we initialize the domains $(t_2, 0)$, where $t_2 = init,init+(N_{skip}+1),\ldots,t = (N_{skip}+1)\times i_t+init \le NT-1$, and $i_t=0,1,\ldots,\ceil{\frac{NT}{(N_{skip}+1)}}$.
Here, $init = 0$ if $N_{skip} = 0$, $init = 1$ if $N_{tskip} = 1$, or $init = 2$ if $N_{skip} = 2$. Again, the same $NP_{bf}$ and $NG_{bf}$ are used from the backward information propagation. The processing and interpolation are then performed similarly to the previous steps, keeping the results with a higher coherence value when compared to the ones obtained previously. The execution ends after the domains $(t_2, Ntr - 1)$ are processed, generating a set of optimal parameters for each domain within an interactive domain.
The backward-and-forward propagation is repeated for as many $N_{cycle}$ as defined by the user.

Finally, this work uses the JADE algorithm with archives as a search engine within the domains, configuring as previously described.
Besides, we suggest $NP_{bf}>4$ so that the method is always looking for new random individuals to avoid any selection problem, which can prevent the metaheuristic from converging to a global optimum. 
Furthermore, we highlight that our implementation is not final, and there is plenty of room to optimize the ENS architecture.

\subsubsection{ENS Semblance computations}

Similarly to JADE, we calculate how many times the ENS algorithm performs Semblance function calls.
The parameters $NP_{f}$, $NG_{f}$, $NP_{bf}$, $NG_{bf}$, $N_{cycle}$, and $N_{skip}$ are the same as described in the ``Evolution by neighborhood similarity'' section. Also, $P_d$ represents all sample points in the output dataset.

In the forward-exploration propagation, $NP_{f}$ individuals are initialized once and then mutated during $NG_{f}$ generations per sample. Therefore, it performs
\begin{equation}
 S_{f} = NP_{f}\times\left(1\ +\ NG_{f}\right)
\end{equation}
Semblance operations per sample.

On the $N_{cycle}$ propagation pairs, we initialize only $\left(NP_{bf}-1\right)$ individuals, as one of the population members already has its Semblance value calculated from the previous propagation; those $NP_{bf}$ individuals then evolve for $NG_{bf}$ generations, thus performing
\begin{equation}
S_{bf} = 2 \times \left(NP_{bf}\times\left(1+NG_{bf}\right)-1\right)
\end{equation}
Semblance computations per sample.

Furthermore, each skipped sample performs an interpolation and one more Semblance computation. Therefore, we compute $N_{skip}$ more Semblance operations for each estimated sample. Adding up all $\ceil{\frac{P_d}{N_{skip}+1}}$ estimated samples, we have that the total number of Semblance operations performed is 
\begin{eqnarray}
\label{eqn:ens_semblance}
num\_ens\_semblances\_calls &=& \ceil{\frac{P_d}{N_{skip}+1}} \nonumber \\
&\times& \left(S_{f} + N_{cycle} \times S_{bf}\right. \nonumber \\
&+&\left. (2 \times N_{cycle} + 1) \times N_{skip} \right)\, .
\end{eqnarray}

\section{Computational aspects}

Given a coordinate system, we call the process of making the trace grid of the seismic-response dataset uniform of regularization. Due to problems in dataset acquisition, the dataset grid often comes with missing or muted traces, irregular trace positions in a specific geometry, and incomplete binning. Therefore, regularization is necessary so that the dataset is in the ideal format for the following seismic processing and imaging stages. The traveltime parameter searching is performed on the sample traces on that new grid. On the other hand, the process does not create a new grid when the grid is already regular. It only estimates the parameters in that grid and is usually used to improve the signal-to-noise ratio. In computational terms, the regularization process has four main stages: grid geometry preparation, bucket generation, bucket processing, and bucket committing. Those are all coordinated by a task system.

In order to clarify, we highlight that we define a bucket as a subset of traces in the dataset that generate the ENS interactive domain. The choice to split the dataset into multiple buckets originates from memory limitations, especially on GPUs, making it not viable to perform the process on the entire dataset at once. Furthermore, as ENS uses neighbor information, the traces in a bucket must have similar characteristics. The first step of the regularization process, grid geometry preparation, sorts the data in order to satisfy such requirements.

The next step, bucket generation, involves selecting the $N_{bucket}$ traces in a bucket and sending them for processing. The user can configure the number of traces $N_{bucket}$, which is a relevant parameter to guarantee the best quality. Thus, the bucket processing step is where the actual ENS execution happens, and we obtain the best parameters set for each trace. The results from separate buckets are then joined in the bucket committing step. Besides, $N_{bucket}$ is often different from $Ntr$ due to physical memory limitations.

\subsection{The scalable partially idempotent task
system}\label{the-scalable-partially-idempotent-task-system}

\begin{figure}[thb!]
\begin{center}
\includegraphics[width=0.84\columnwidth]{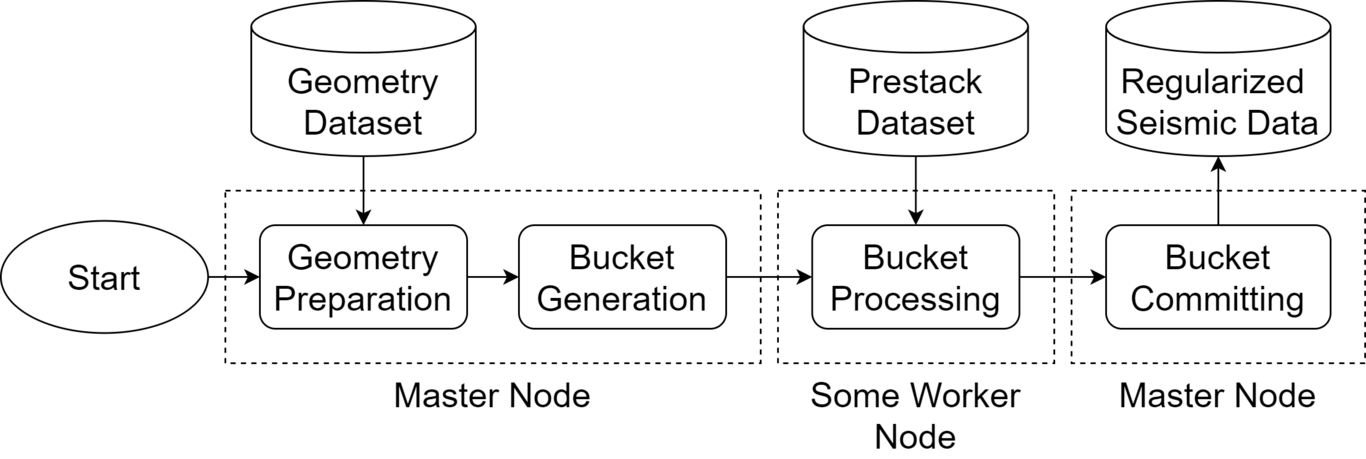}
\caption{{{Complete Dataset regularization involving the four main stages of processing and their data dependencies}.
{\label{fig:regularization_flowchart}}%
}}
\end{center}
\end{figure}

The Scalable Partially Idempotent Task System (SPITS) \citep[see,][]{pypits} is a task system that provides fault tolerance by default, meaning that we can run our code over a distributed unreliable cluster of machines without worrying that one instance fault may take down the whole execution. Another advantage of SPITS is its ability to gracefully scale the execution by simply adding or removing instances from the system. Finally, the embarrassingly parallel property of SPITS is ideally suited to our parameter estimation process since the regularization process can be partitioned into multiple independent estimations. Figure~\ref{fig:regularization_flowchart} exemplifies how our system takes advantage of the SPITS computational paradigm.

In summary, we build the previously described four steps on top of SPITS. First, the master node loads the geometry dataset as the grid geometry preparation. Next, in the bucket generation step, the master node splits the received dataset into multiple buckets. It sends them to the workers in the system using a round-robin fashion \citep{Miao2016}. Besides, the worker nodes' estimation process to process their buckets can proceed either in the CPU or GPU. We advise the usage of GPUs to power the execution due to the highly independent nature of each time sample computation. Finally, after the estimation process is finished, the estimated results are sent back to the master node to be committed into a single final file in the bucket committing step.

\section{Numerical experiments}

Although the ENS technique is agnostic regarding CPU or GPU usage, we have chosen to proceed with an ENS with GPU implementation (ENS-GPU), as this kind of hardware is more suitable to solve the traveltime parameter search problem. Similarly, we have also implemented a JADE with GPU (JADE-GPU) to compare against ENS-GPU. We highlight that when comparing ENS and JADE sans the -GPU suffix, we are comparing the techniques, not the specific implementation. Such comparisons will be more common in the qualitative analysis subsection, while performance-wise, the GPU implementation will be focused. Besides, in~\cite{Ribeiro2019,Ribeiro2020}, the JADE-GPU/CPU approach performed better than the DE-GPU/CPU approach, which is why it was chosen to challenge our approach.

We considered three different datasets to validate the ENS-GPU approach (Table~{\ref{tab:datasets}}). The geometry used for the regularization is the same as the prestack configuration. Dataset one is an inline synthetic data modeled by a Kirchhoff integral algorithm (Dataset~1), dataset two is a land prestack extracted from the Tacutu Basin, located in the northeast region of Roraima, Brazil (Dataset~2), and the third dataset is a marine 2D line acquisition from the coast of Bahia, Brazil (Dataset~3). Alongside the type of data, we depict in each entry the overall data size, the number of seismic traces, and finally, the total number of time samples of each specific dataset. In these experiments, we considered the FO-CRS and OCT, in which for both, we used a midpoint aperture of 100m, an offset aperture of 2000m, and a Semblance window of 48ms with all the estimations. Important to note that the sample interval of all tested datasets is 4ms. Furthermore, Table~\ref{tab:moved_traces} shows the average number of traces moved from the host (CPU) memory to the device (GPU) memory per seismic trace estimation and for each traveltime. We also show in this table the average number of traces used during the Semblance computation since this value varies according to the specified time sample being estimated, especially for OCT due to its nature, where the difference between the number of traces moved and the number of traces effectively used by Semblance can be almost twenty times.

% \begin{table}[thb!]
% \centering
% \normalsize
% \normalsize\begin{tabular}{|c|c|c|c|c|c|c|c|}
% \hline
% \textbf{Geometry Dataset} & \textbf{Size (MB)} & \textbf{Number of Traces ($Ntr$)} & \textbf{Time Samples ($NT$)} & \textbf{Moved Traces (FO-CRS)} & \textbf{Used Traces (FO-CRS)} & \textbf{Moved Traces (OCT)} & \textbf{Used Traces (OCT)} \\ \hline
% \textbf{Dataset~1} & 4 & 2626 & 376 & 280 & 210 & 2094 & 225 \\ \hline
% \textbf{Dataset~2} & 75 & 18048 & 1024 & 89 & 50 & 1014 & 53 \\ \hline
% \textbf{Dataset~3} & 402 & 58189 & 1751 & 495 & 375 & 5091 & 390 \\ \hline
% \end{tabular}
% \caption{{{Seismic survey datasets used in the experiments.}
% {\label{tab:datasets}}%
% }}
% \end{table}
\begin{table}[thb!]
\centering
\adjustbox{max width=\textwidth}{\begin{tabular}{|c|c|c|c|}
\hline
\textbf{\makecell{Geometry\\ Dataset}} & \textbf{\makecell{Size\\ (MB)}} & \textbf{\makecell{Number of Traces\\ ($Ntr$)}} & \textbf{\makecell{Time Samples\\ ($NT$)}} \\ \hline
\textbf{Dataset~1} & 4 & 2626 & 376 \\ \hline
\textbf{Dataset~2} & 75 & 18048 & 1024 \\ \hline
\textbf{Dataset~3} & 402 & 58189 & 175 \\ \hline
\end{tabular}}
\caption{{{Seismic survey datasets used in the experiments.}
{\label{tab:datasets}}%
}}
\end{table}
\begin{table}[thb!]
\centering
\adjustbox{max width=\textwidth}{\begin{tabular}{|c|c|c|c|c|}
\hline
\textbf{\makecell{Geometry\\ Dataset}} & \textbf{\makecell{Moved Traces\\ (FO-CRS)}} & \textbf{\makecell{\textbf Used Traces\\ (FO-CRS)}} & \textbf{\makecell{\textbf Moved Traces\\ (OCT)}} & \textbf{\makecell{\textbf Used Traces\\ (OCT)}} \\ \hline
\textbf{Dataset~1} & 280 & 210 & 2094 & 225 \\ \hline
\textbf{Dataset~2} & 89 & 50 & 1014 & 53 \\ \hline
\textbf{Dataset~3} & 495 & 375 & 5091 & 390 \\ \hline
\end{tabular}}
\caption{{{Numbers of traces moved and used for semblance computation in the experiments.}
{\label{tab:moved_traces}}%
}}
\end{table}

To process these datasets, we used one computing node powered by a dual-socket Intel Xeon Gold 6148 CPU, 188GB of RAM, and four Nvidia Tesla V100 with 16GB of memory each. The regularization code was implemented in \textbf{C++} and~\textbf{CUDA}, compiled with \textbf{GCC} 5.4.0 and \textbf{NVCC} 9.0.176 using the \textbf{-O3} optimization flag. We use the SPITS programming model and the \textbf{PY-PITS} framework by \citet{pypits} for the orchestration and parallel execution of the buckets. Furthermore, for all experiments, we fix the ENS configuration parameters $NP_{bf} = 5$, $NG_{bf} = 1$, $N_{skip}=2$, and $N_{cycle}=1$. The execution of time samples, the $t_i$, where $i=0,1,2$, described in the evolution by neighborhood similarity section, is done in parallel in the GPU.

Initially, we used only one worker to extract the full potential of the four GPUs in our node. The worker was responsible for processing a bucket of 640 traces in the OCT estimations and a bucket of 2560 traces in the FO-CRS estimation. Furthermore, virtual threads are introduced to the computing worker to reduce idle time on GPU. The received bucket is split between the virtual threads, then processed in parallel using the same GPU. Note that GPU idleness is reduced at the cost of memory usage increase, as multiple threads are now loading traces to GPU simultaneously. Finally, due to that memory constraint, we use a total of forty virtual threads (ten per GPU), and we split the bucket into smaller buckets so that each one of the forty traces can process the same amount of traces (16 traces per thread for OCT and 64 traces per thread for FO-CRS). It is relevant to state that those configuration settings are specific for the GPU implementation and do not translate to a CPU implementation. For instance, the CPU implementation does not suffer from the same data movement overhead, as the data is readily available for usage. For the CPU-specific configurations, another article would be necessary, although we still expect the ENS-CPU to be considerably faster than JADE-CPU due to the convergence speed of the first technique.

\subsection{Qualitative analysis}

Given that this part is highly based on empirical, subjective, and visual interpretation information, we start presenting our results with the Semblance panels.
All results are obtained without considering any initial traveltime parameter to guide the processes, making the comparison between strategies fairer.
Figure~{\ref{fig:focrs_semblance_dataset1}} shows the FO-CRS Semblance results obtained after the parameter estimation of one common-offset panel from Dataset~1.
Similarly, Figure~{\ref{fig:oct_semblance_dataset1}} presents the OCT Semblance results related to Dataset~1. The results are presented in a grid view, where the first line has values generated by the JADE metaheuristic, whereas our ENS strategy generates the second line. Also, the first column, the leftmost one, is related to the execution of an evolutionary process with $NP_f=NP=5$ and $NG_f=NG=1$. Similarly, the second column is related to $NP_f=NP=30$ and $NG_f=NG=30$. Notably, both JADE and ENS generated very similar results when compared in the second column. However, JADE produced unacceptable results for the first column configuration. In comparison, ENS generated acceptable results similar to those obtained after an increase from $(NP_f=5, NG_f=1)$ to $(NP_f=NG_f=30)$. However, ENS launches three mutation steps (Forward, Backward-and-Forward propagations), unlike JADE. If we plug in $NP_f=NP=5$ and $NG_f=NG=1$ in eqs~({\ref{eq:jade_semblance}}) and~({\ref{eqn:ens_semblance}}), we highlight that they perform the same amount of Semblance computations for both approaches, i.e., ten Semblance computations per sample for JADE and a little bit more than eleven Semblance computations per sample for ENS. Furthermore, the same pattern repeats with the second and third dataset results presented in figures~{\ref{fig:focrs_semblance_dataset2}},~{\ref{fig:oct_semblance_dataset2}},~{\ref{fig:focrs_semblance_dataset3}}, and~{\ref{fig:oct_semblance_dataset3}}. Although for FO-CRS, the quality is significantly reduced when using JADE. We observe that these qualitative results reflect the stacked results presented in the figures~{\ref{fig:focrs_stacking_dataset2}},~{\ref{fig:oct_stacking_dataset2}},~{\ref{fig:focrs_stacking_dataset3}}, and~{\ref{fig:oct_stacking_dataset3}}. Finally, in the FO-CRS case, even with $(30\times 30)$, JADE does not return a result with the same quality as ENS with $(5\times 1)$. While for the OCT case, JADE manages to get closer to the ENS, but still in an inferior way. Furthermore, pay attention that the ratio $N_{bucket}(\text{CRS})/N_{bucket}(\text{OCT})$ is four, which can increase by up to four times the amount of similar domains to FO-CRS concerning OCT.

\newcommand\qualitativespace{0.8}

\begin{figure}[thb!]
\begin{center}
\includegraphics[width=\qualitativespace\columnwidth]{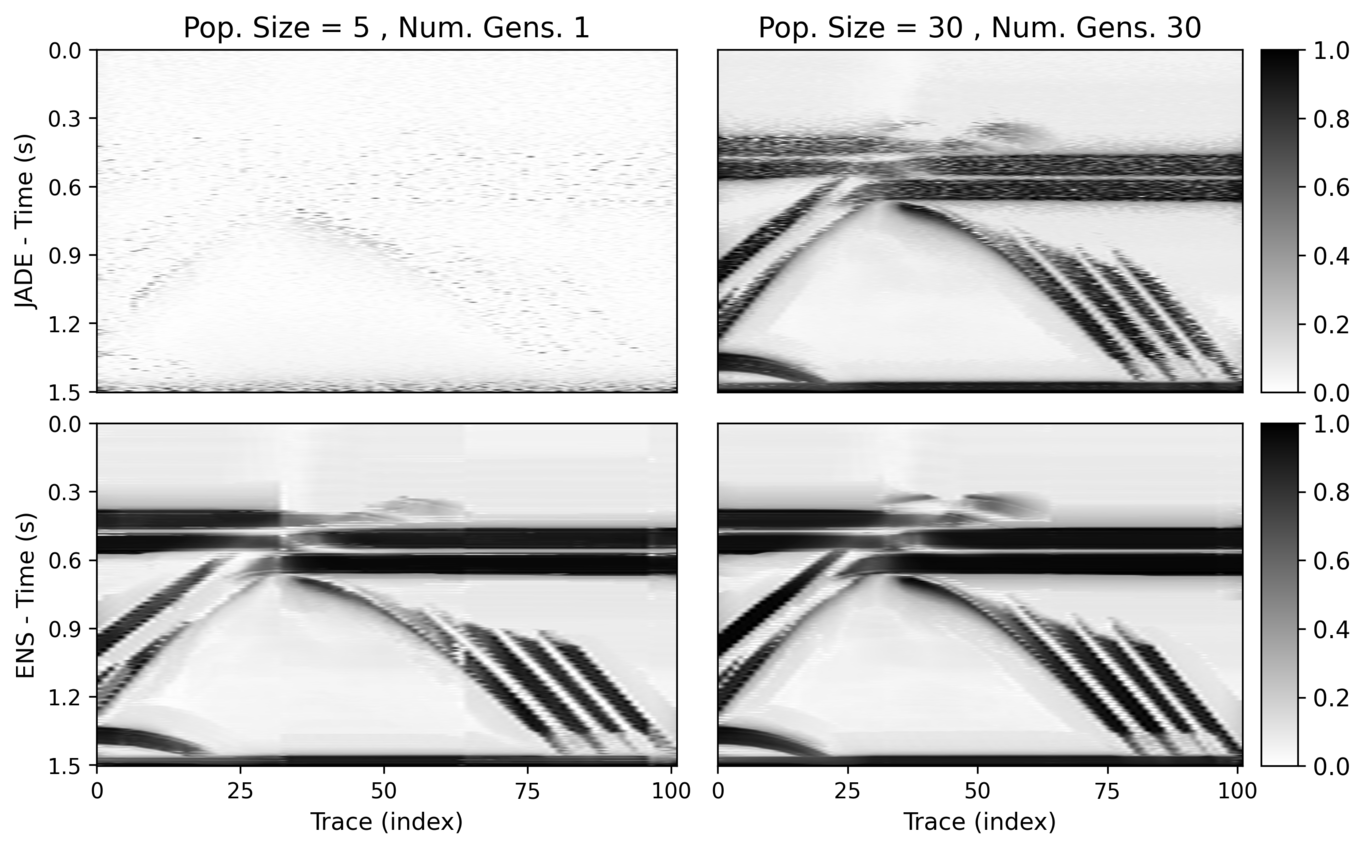}
\caption{{{FO-CRS Semblance panels of common offset 500m of Dataset~1 (Bucket size = 64).}
{\label{fig:focrs_semblance_dataset1}}%
}}
\end{center}
\end{figure}
\begin{figure}[thb!]
\begin{center}
\includegraphics[width=\qualitativespace\columnwidth]{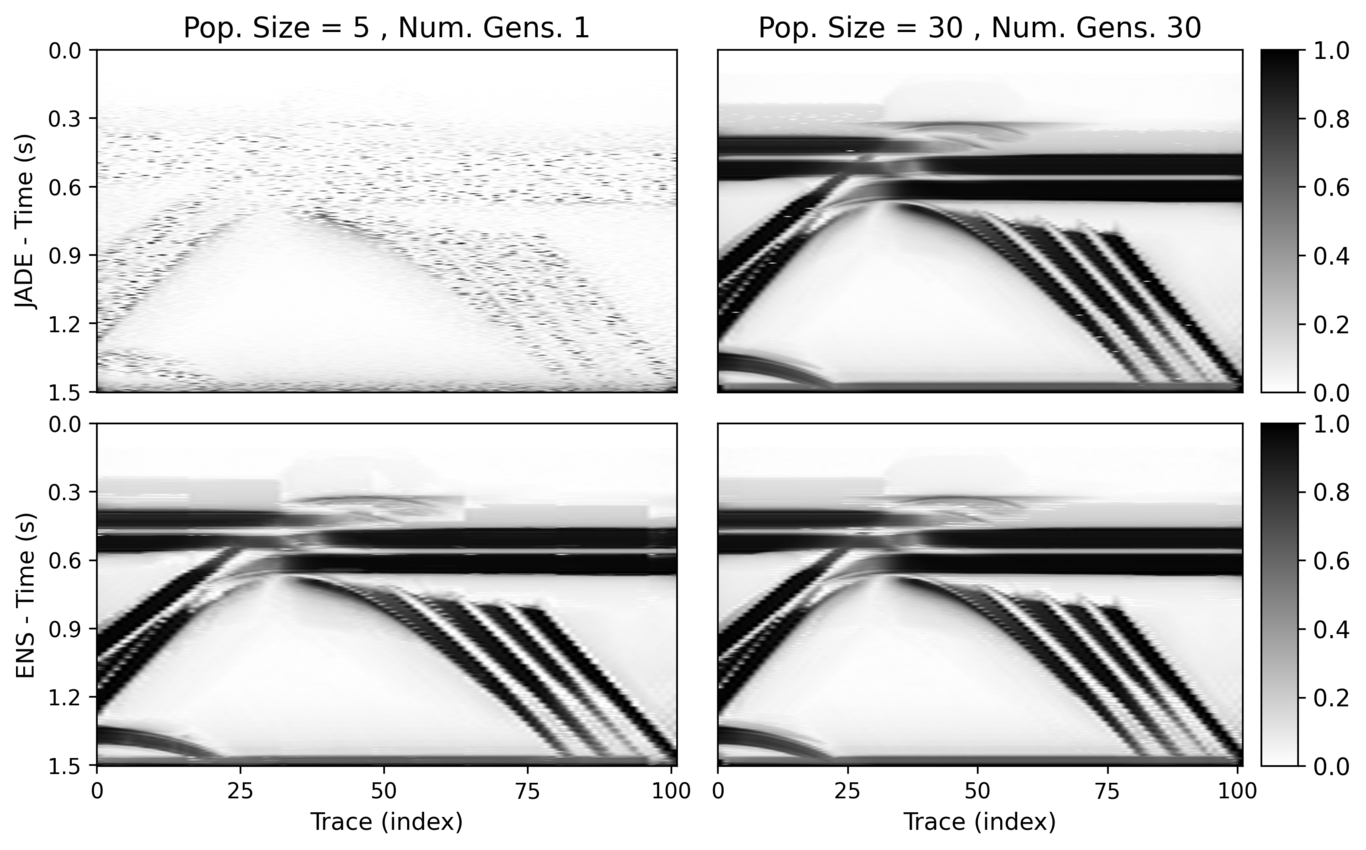}
\caption{{{ OCT Semblance panels of common offset 500m of Dataset~1 (Bucket size = 16).}
{\label{fig:oct_semblance_dataset1}}%
}}
\end{center}
\end{figure}
\begin{figure}[thb!]
\begin{center}
\includegraphics[width=\qualitativespace\columnwidth]{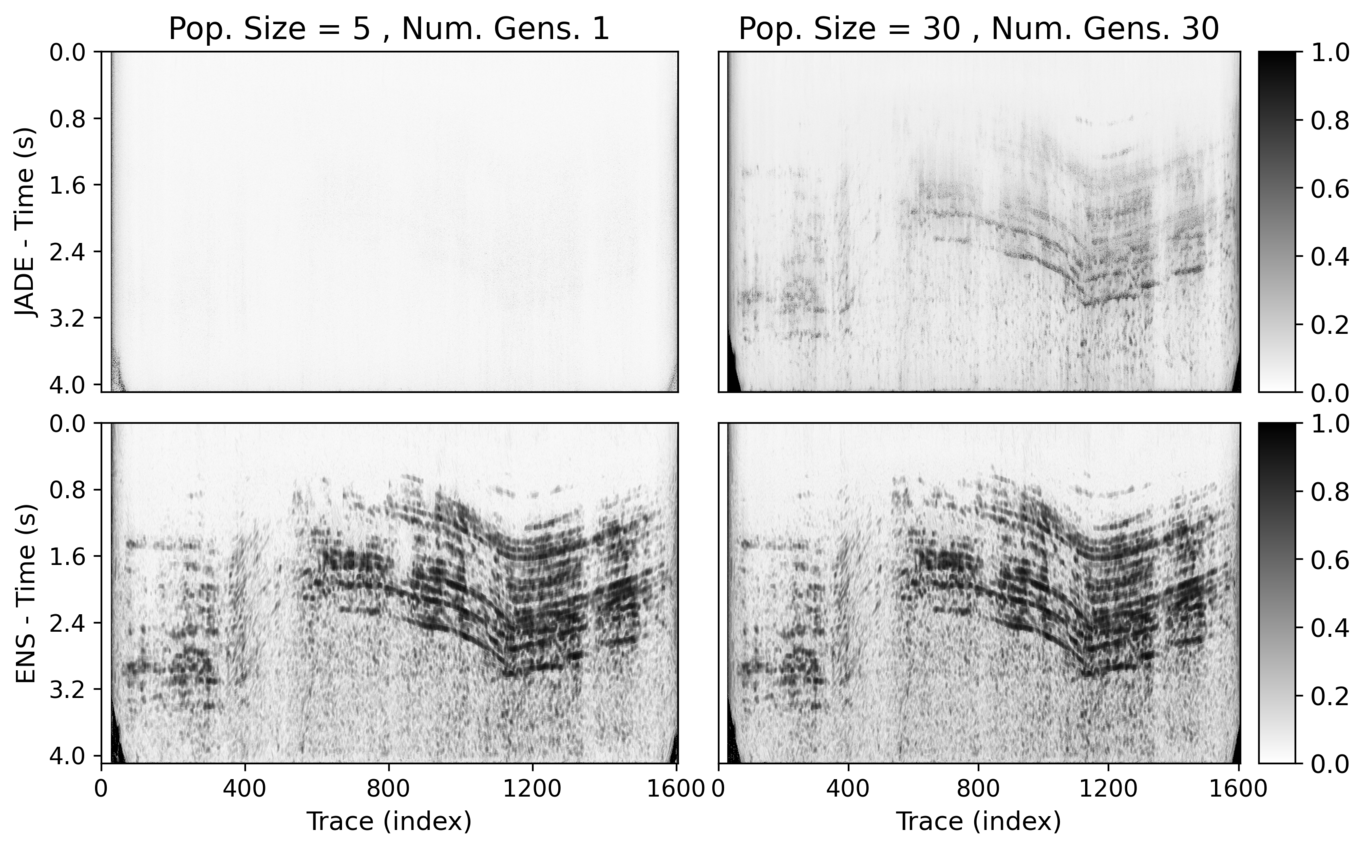}
\caption{{{ FO-CRS Semblance panels of common offset 1000m of Dataset~2 (Bucket size = 64).}
{\label{fig:focrs_semblance_dataset2}}%
}}
\end{center}
\end{figure}
\begin{figure}[thb!]
\begin{center}
\includegraphics[width=\qualitativespace\columnwidth]{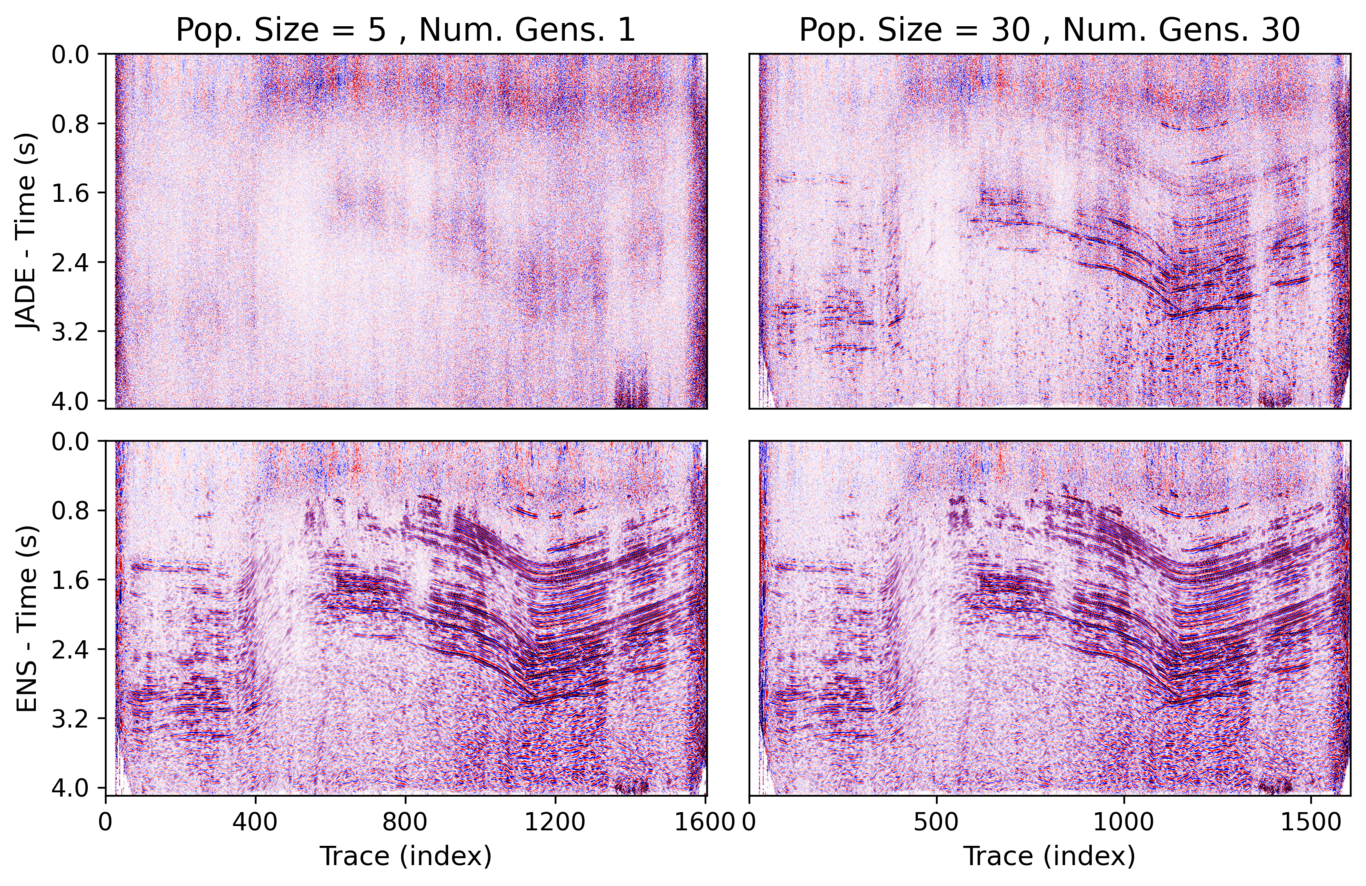}
\caption{{{ FO-CRS stacking panels of common offset 1000m of Dataset~2 (Bucket size = 64).}
{\label{fig:focrs_stacking_dataset2}}%
}}
\end{center}
\end{figure}
\begin{figure}[thb!]
\begin{center}
\includegraphics[width=\qualitativespace\columnwidth]{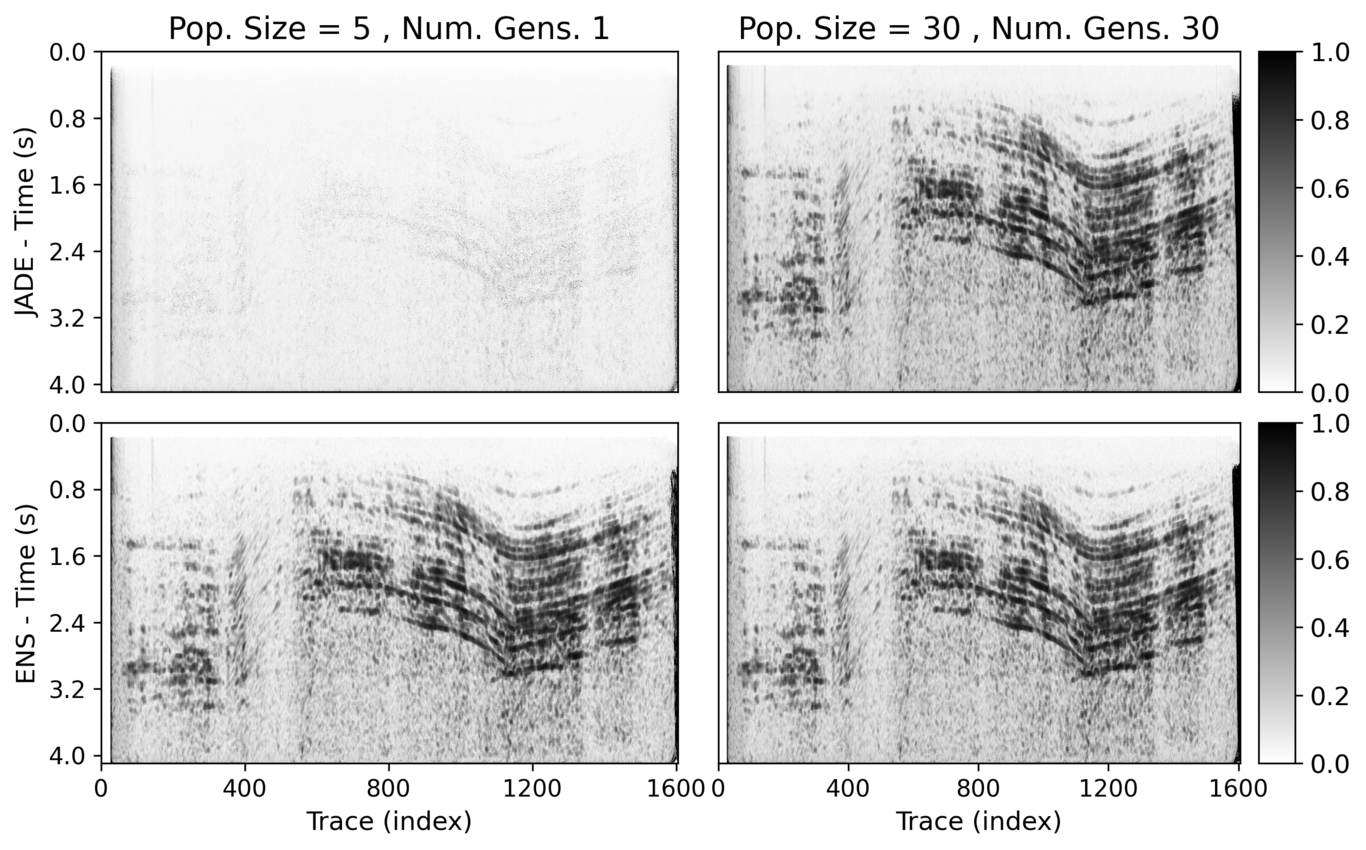}
\caption{{OCT Semblance panels of common offset 1000m of Dataset~2 (Bucket size = 16).
{\label{fig:oct_semblance_dataset2}}%
}}
\end{center}
\end{figure}
\begin{figure}[thb!]
\begin{center}
\includegraphics[width=\qualitativespace\columnwidth]{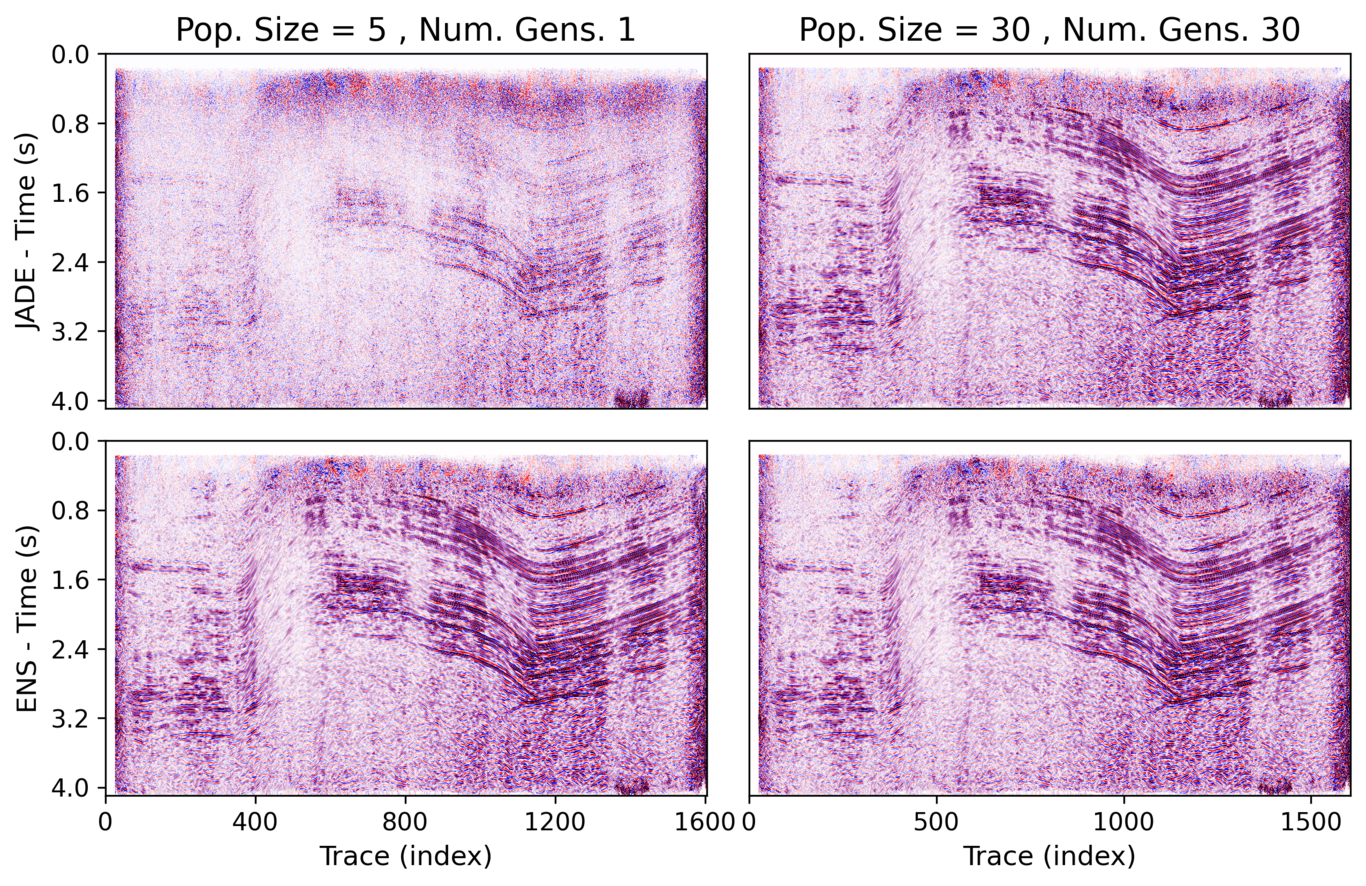}
\caption{{OCT stacking panels of common offset 1000m of Dataset~2 (Bucket size = 16).
{\label{fig:oct_stacking_dataset2}}%
}}
\end{center}
\end{figure}
\begin{figure}[thb!]
\begin{center}
\includegraphics[width=\qualitativespace\columnwidth]{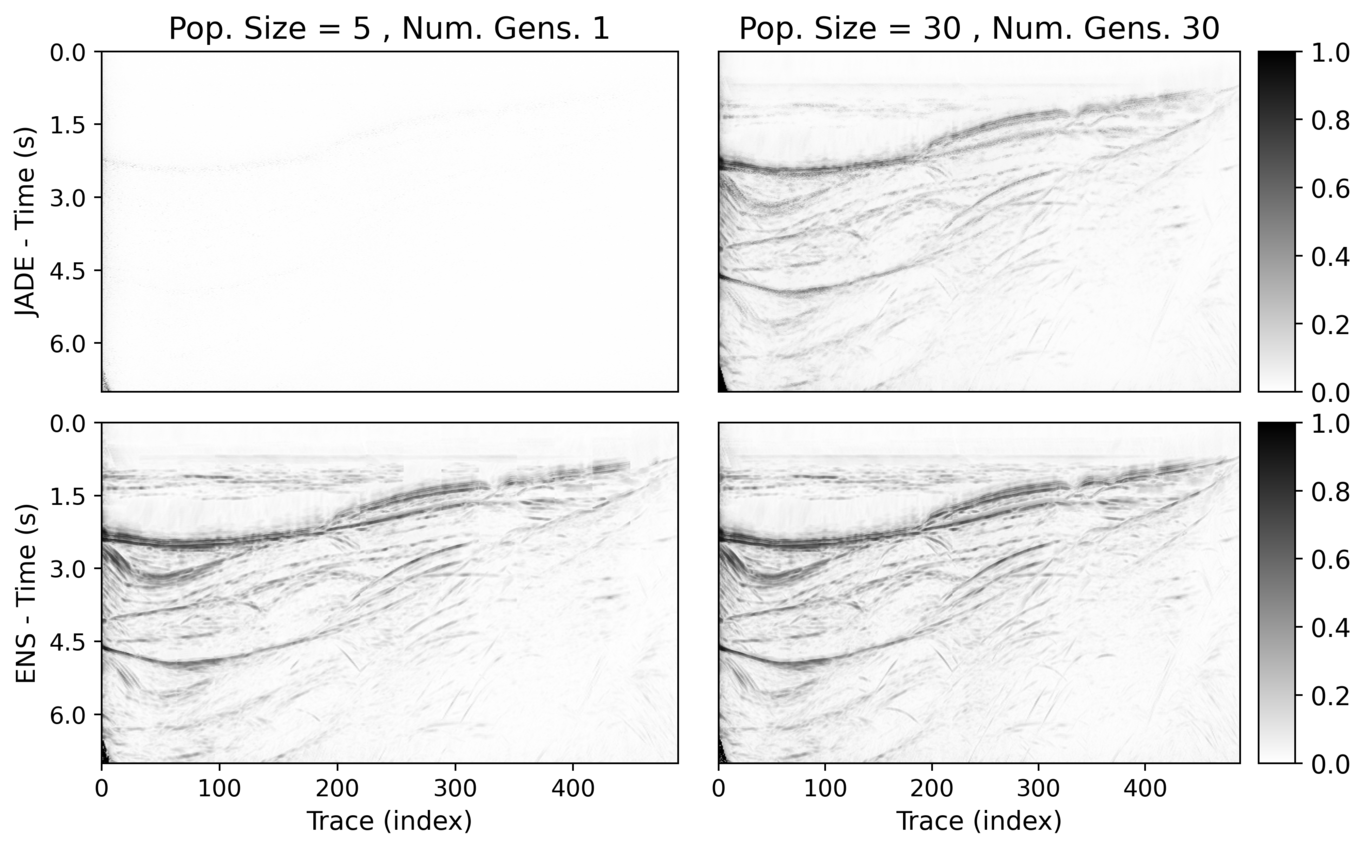}
\caption{{{FO-CRS Semblance panels of common offset 1000m of Dataset~3 (Bucket size = 64).}
{\label{fig:focrs_semblance_dataset3}}%
}}
\end{center}
\end{figure}
\begin{figure}[thb!]
\begin{center}
\includegraphics[width=\qualitativespace\columnwidth]{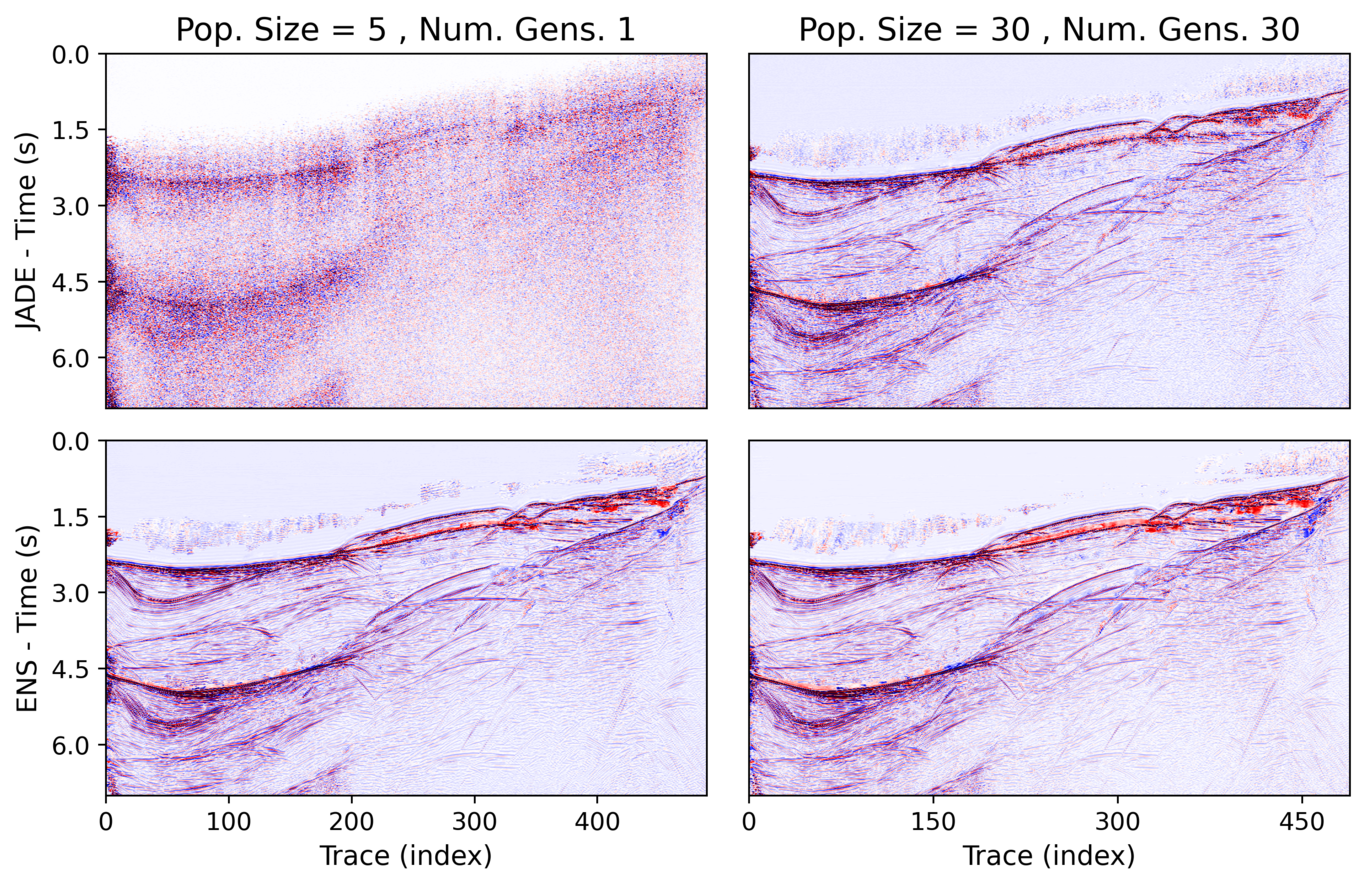}
\caption{{{FO-CRS stacking panels of common offset 1000m of Dataset~3 (Bucket size = 64).}
{\label{fig:focrs_stacking_dataset3}}%
}}
\end{center}
\end{figure}
\begin{figure}[thb!]
\begin{center}
\includegraphics[width=\qualitativespace\columnwidth]{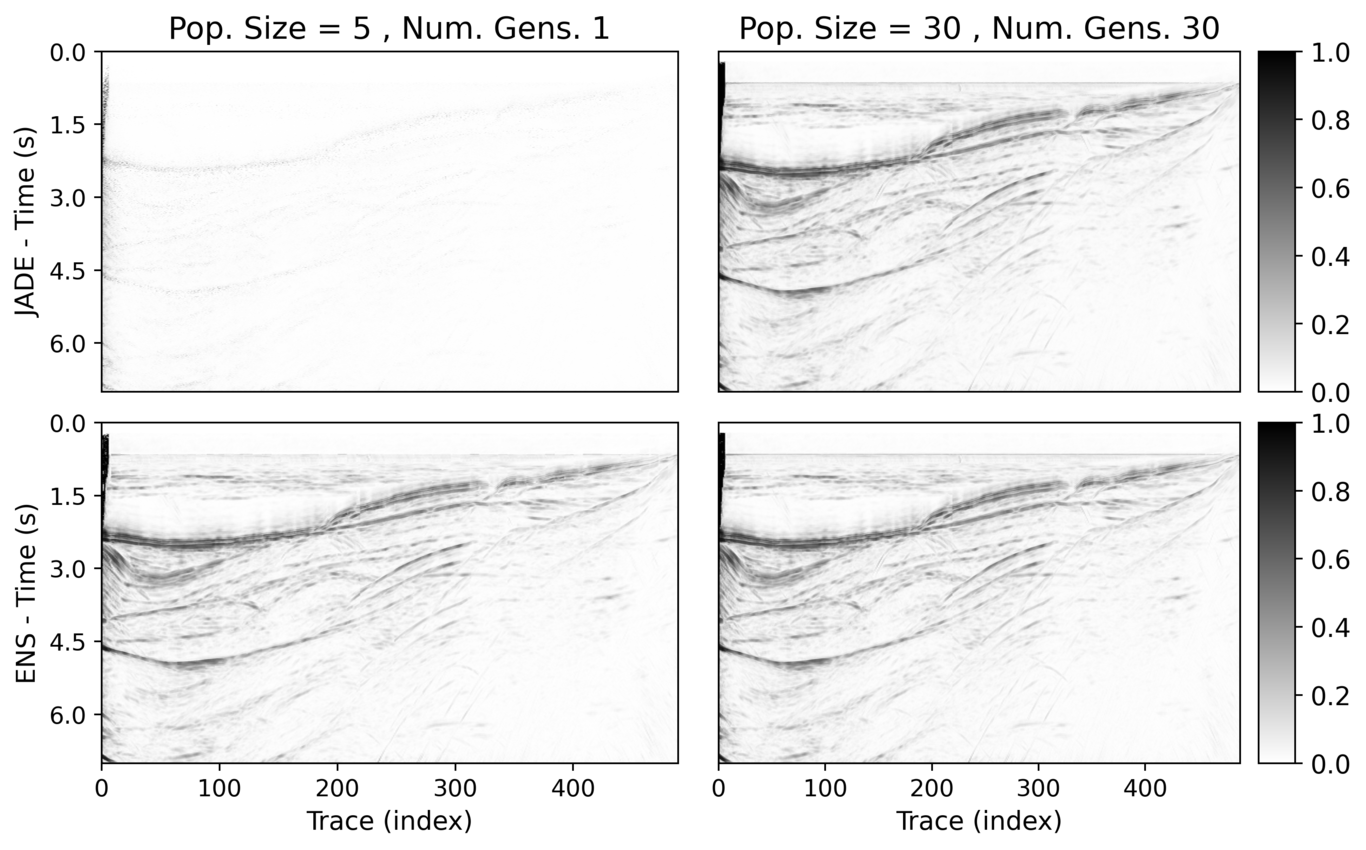}
\caption{{OCT Semblance panels of common offset 1000m of Dataset~3 (Bucket size = 16).
{\label{fig:oct_semblance_dataset3}}%
}}
\end{center}
\end{figure}
\begin{figure}[thb!]
\begin{center}
\includegraphics[width=\qualitativespace\columnwidth]{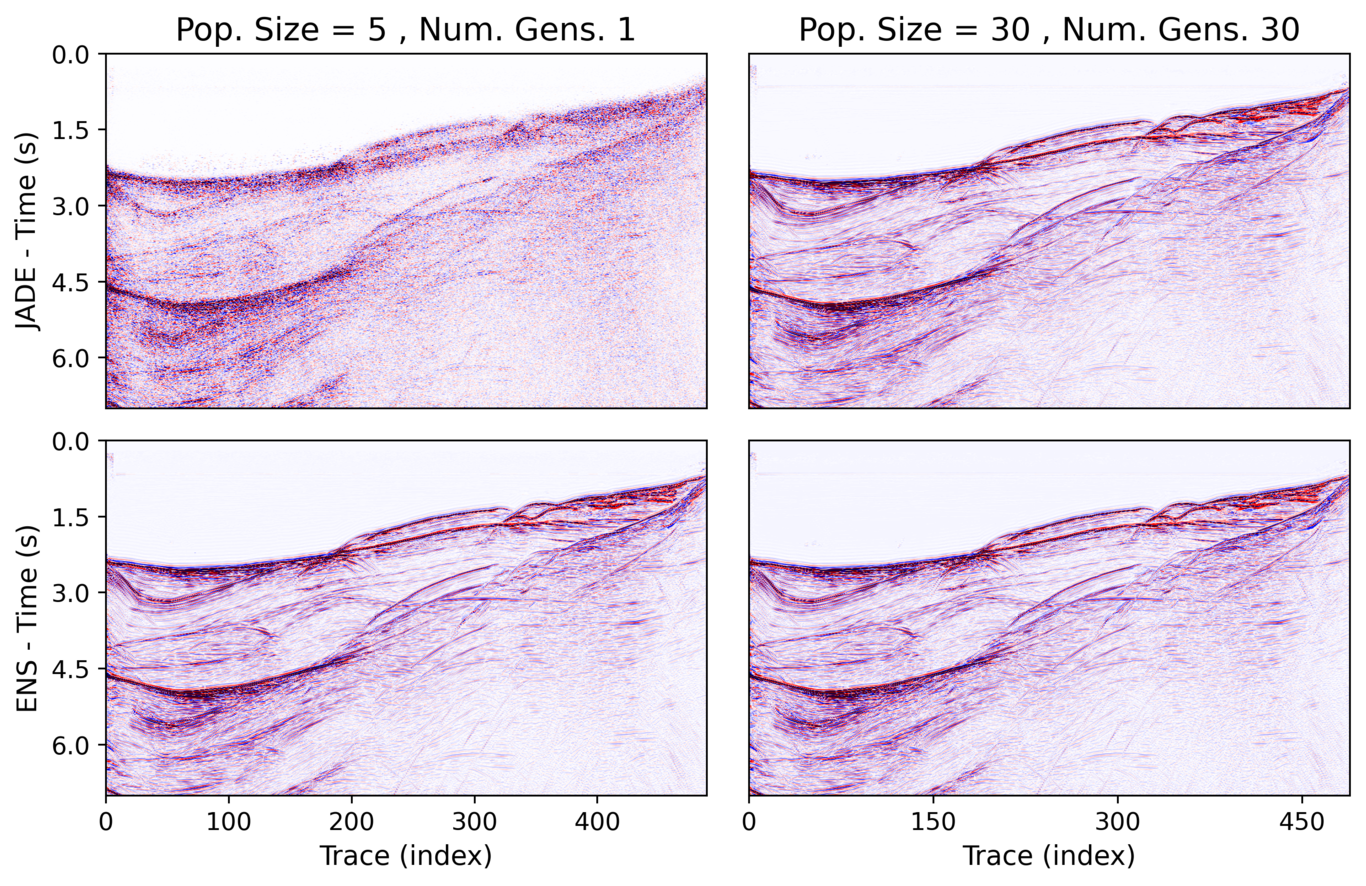}
\caption{{OCT stacking panels of common offset 1000m of Dataset~3 (Bucket size = 16).
{\label{fig:oct_stacking_dataset3}}%
}}
\end{center}
\end{figure}

\subsection{Quantitative analysis}

\begin{figure}[thb!]
\begin{center}
\includegraphics[width=1.00\columnwidth]{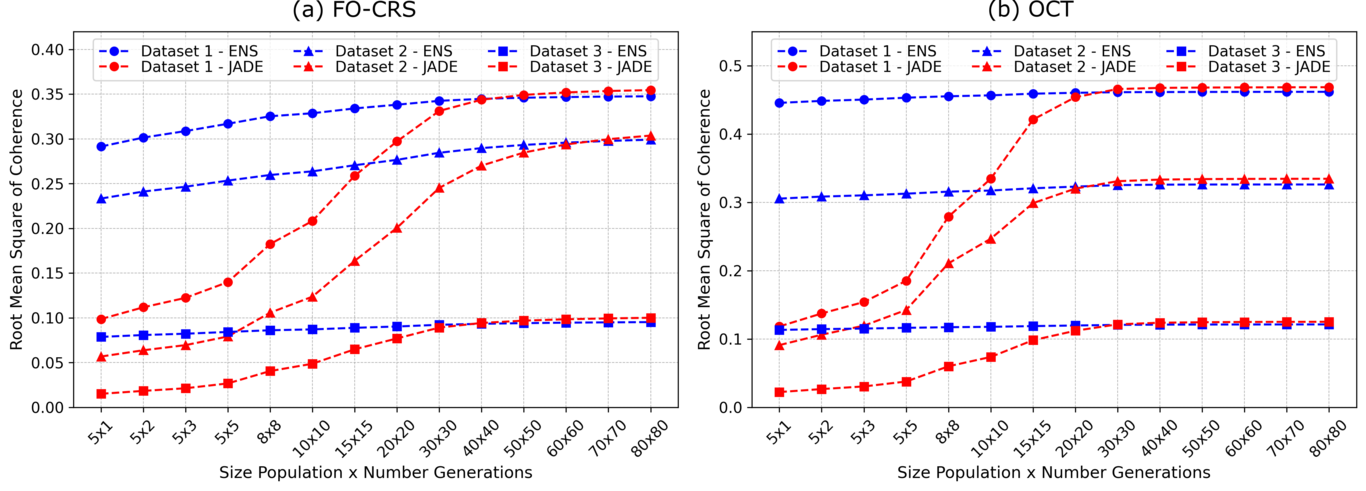}
\caption{{Coherence analysis - ENS x JADE (FO-CRS Bucket size = 64, OCT Bucket size = 16).
{\label{fig:rms_coherence}}%
}}
\end{center}
\end{figure}

The qualitative properties observed in the previous Semblance panels are extensible to numerical analysis. For instance, Figure~{\ref{fig:rms_coherence}} compares the root mean square (RMS) value obtained from the coherence panels in each dataset (the $g$-type fitness function here is the RMS). In this comparison, JADE and ENS are executed in multiple evolutionary settings, where the population size and the number of generations vary (here, $NP_f=NP$ and $NG_f=NG$). These results show that the RMS of coherence panels produced by ENS (blue lines) has not significantly improved when the settings go from a population with five individuals mutated through 1 generation to a population with 80 individuals mutated through 80 generations. On the other hand, panels obtained out of the pure JADE execution (red lines) depict a very noticeable improvement when the first setting ($5\times 1$) is compared with the last one ($80\times 80$). The direct impact of these results implies that JADE requires many more computations to converge compared to ENS. Although the more complex algorithm of ENS makes its iteration slower than JADE, a speedup analysis is required to confirm if the smaller number of generations and population size also produce faster execution times. However, despite the slow convergence, JADE achieves a higher coherence RMS than the current implementation ENS in more significant populations/generations. As $N_{skip}= 2$, the performance-quality trade-off by interpolation causes such a difference. Finally, the results obtained with the execution of the FO-CRS (Figure~\ref{fig:rms_coherence}~(a)) have a much lower quality than those obtained by the OCT method (Figure~\ref{fig:rms_coherence}~(b)). Furthermore, given the number of traveltime parameters, the OCT is more robust than FO-CRS as the number of searches decreases in ENS. 

\subsection{Performance results}

\begin{figure}[thb!]
\begin{center}
\includegraphics[width=1.00\columnwidth]{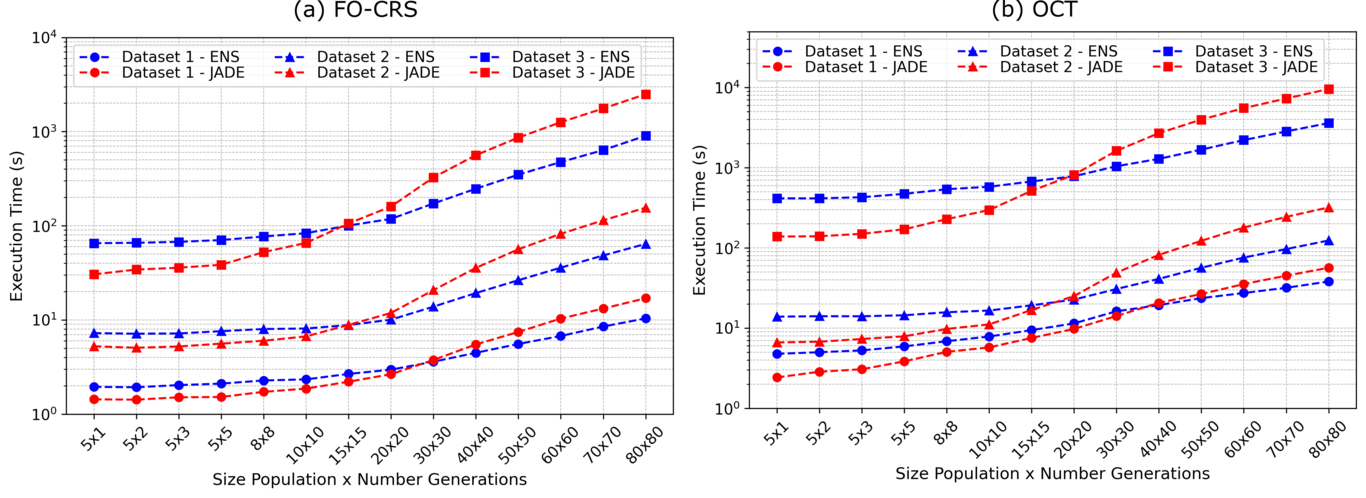}
\caption{{Execution Time - ENS-GPU x JADE-GPU (FO-CRS Bucket = 64, OCT Bucket = 16, number GPUs = 4, threads per GPU = 10).
{\label{fig:ens_jade_execution_time}}%
}}
\end{center}
\end{figure}

To verify how this fast convergence speed of ENS-GPU translates in terms of execution time, we present Figure~\ref{fig:ens_jade_execution_time}, which are the absolute and relative execution times, respectively, for the (a) FO-CRS and~(b) OCT time results of both JADE-GPU and ENS-GPU for multiple datasets. Note that we use the suffix ``-GPU'' to refer to our ENS and JADE algorithm implementations on GPUs, while not using the suffix indicates the theoretical side of the techniques.

Contrarily to the desired scenario, in Figure~\ref{fig:ens_jade_execution_time}~(b), the ENS-GPU of an OCT execution is considerably more expensive than JADE-GPU on small populations and fewer generations, being almost three times slower in Dataset~3, the biggest one. However, this undesired scenario changes after the $(20\times 20)$ configuration, such that JADE becomes more expensive than ENS, taking twice as long to perform the regularization in the third dataset using an $(80\times 80)$ configuration. Furthermore, this pattern repeats for Dataset~3 in the FO-CRS executions from Figure~\ref{fig:ens_jade_execution_time}~(a), though earlier than expected. Speedup inverts when the settings are still $(15\times 15)$, making ENS-GPU more attractive than JADE-GPU in an earlier setting compared to OCT.

Also, it is noticeable that Dataset~3 presents the most extreme results. For instance, when the settings are small ($5\times 1$, $5\times 2$), the speedup results of ENS-GPU over JADE-GPU are the worst out of all the other datasets. On the other hand, when the settings are more extensive ($70\times 70$, $80\times 80$), the speedup results of ENS-GPU over JADE-GPU are the best of all executions. Since Dataset~3 is the biggest one (more than three times bigger than the second biggest), this certainly raises the issue that this speedup behavior may be related to data overhead. After all, data overhead is expected to significantly impact the execution time when computations are fast (smaller settings), while it is expected to be negligible when computations are expensive (more significant settings).

We highlight that the number of Semblance evaluations to obtain the JADE and ENS results can be calculated using eqs.~(\ref{eq:jade_semblance}) and~(\ref{eqn:ens_semblance}), respectively. In which $P_d$ can be obtained by multiplying the number of traces and time samples from Table~\ref{tab:datasets} for each dataset, $NP_{f} = NP$ represents the population size and $NG_{f} = NG$ represents the number of generations as shown in the charts. Reminder that we set $NP_{bf} = 5$, $NG_{bf} = 1$, $N_{skip} = 2$, and $N_{cycle} = 1$ for ENS. Also, we can observe that JADE and ENS in the $(5\times1)$ configuration perform a total of $10\times P_d$ and $(34/3)\times P_d$ Semblance calculations, respectively. As $NP \times NG$ increases, JADE performs significantly more semblance operations than ENS, converging to ${N_{skip}+1} = 3$ times more computations. For example, on Dataset~3 ($P_d = 58189 \times 1751 \propto 10^8$) with the $(20 \times 20)$ configuration we perform $4.3\times10^{10}$ Semblance computations when using JADE compared to $1.5\times10^{10}$ Semblance computations with ENS.
Thus, the ENS algorithm can obtain a considerably higher coherence RMS at a smaller number of Semblance computations. However, due to computational physics issues, real-time computation is similar.

\subsection{Virtual thread scalability}

The efficient usage of GPU resources during execution is one of the challenges to be addressed for optimal execution time. For instance, we noted that the time spent doing computations by the GPU was considerably low due to the fast ENS-GPU kernel computations. To alleviate this issue, we introduced multiple threads to the execution, in which each processes a subset of the received bucket in parallel, increasing the GPU thread workload. We benchmark Dataset~3, varying the number of threads between one and forty to explore how they affect the execution time. Only a single Nvidia V100 16GB GPU is used with a bucket of 512 traces for the experiment.

\begin{figure}[thb!]
\begin{center}
\includegraphics[width=1.00\columnwidth]{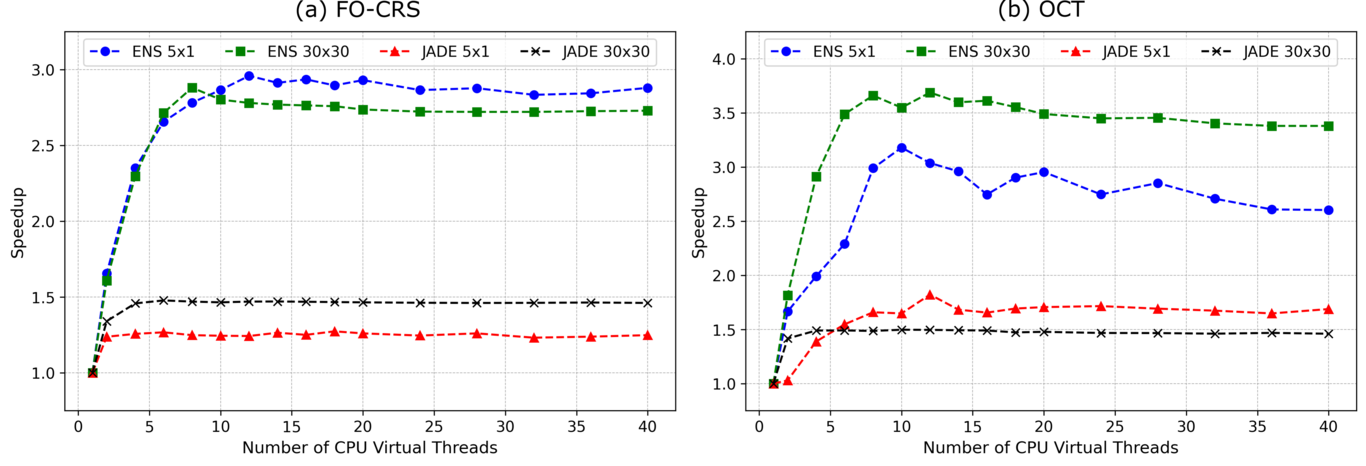}
\caption{{{Scalability of Dataset~3 regarding the increase of CPU virtual
threads. }
{\label{465460}}%
}}
\end{center}
\end{figure}

Figure~\ref{465460}~(a) shows that we obtained almost three times faster executions when using ten threads instead of only one with the FO-CRS using the ENS-GPU application. However, after ten threads, no gains were brought by introducing more threads to execution, thus showing us the limit to the scalability, which is probably related to the number of CUDA cores and streaming multiprocessors available in our GPU. On the other hand, we could obtain only 1.5x speedup with JADE-GPU, so five threads were already sufficient to achieve that speedup. This difference is related to ENS-GPU searches of only one-third of $NT$ samples, while JADE-GPU searches all the $NT$ samples simultaneously. Also, similar results are shown for the OCT execution, Figure~\ref{465460}~(b), where ENS-GPU reaches more than 3.5x speedup by adding ten threads, while JADE-GPU only reaches 2x speedup at maximum.

\begin{figure}[h!]
\begin{center}
\includegraphics[width=1.00\columnwidth]{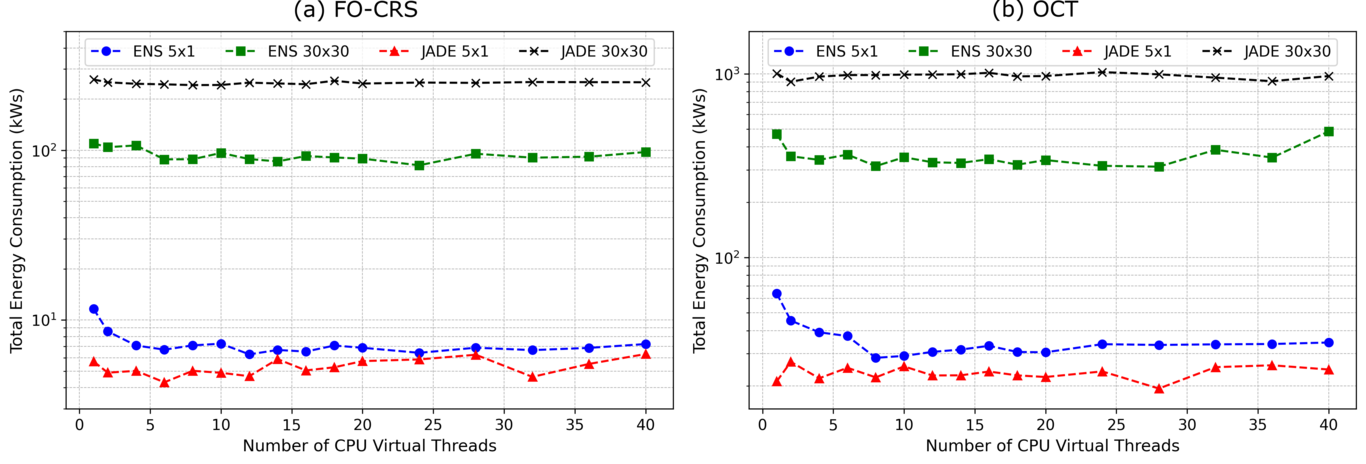}
\caption{{{ GPU total energy consumption in response to a different number of CPU threads.}
{\label{728261}}%
}}
\end{center}
\end{figure}

Another interesting data to consider is the energy consumption by each execution. To analyze it, we measured the average power consumption of the GPU for each varying thread and then multiplied that value by the overall execution time, producing the total energy spent during the whole process. Notably, it is possible to see that, although ENS-GPU had a lower average power consumption than JADE-GPU, the overall energy consumption of ENS-GPU ($5\times 1$) surpasses the JADE-GPU ($5\times 1$) consumption by a slight difference. Despite that, we have shown that an ENS-GPU ($5\times 1$) execution produces acceptable results, while we have to execute a JADE-GPU ($30\times 30$) to obtain similar results to ENS-GPU (Figure~{\ref{fig:rms_coherence}}). However, JADE-GPU ($30\times 30$) with six threads consumes 33 times more energy than ENS-GPU ($5\times 1$) with ten threads, which shows that ENS-GPU is comparatively cheaper than JADE-GPU to execute. Moreover, this is even more highlighted by the OCT results in Figure~\ref{728261}~(b), where a JADE-GPU ($30\times 30$) with ten threads consumes 34 times more energy than an ENS-GPU ($5\times 1$) with ten threads.

However, increasing the number of threads increases memory usage. If we consider that all geometry traces have the same number of traces in their fold, doubling the number of threads implies doubling the memory usage. To some extent, this is acceptable as long as we have enough memory available on GPU. On the other hand, execution may start suffering considerably from the high time overhead generated by the memory extrapolation, as shown by \citet{Okita_2021}.


\subsection{GPU scalability}

GPU Scalability is an essential aspect of the technique to enable the execution of larger datasets in a viable time. Here, we show how ENS-GPU scales when performing the OCT and FO-CRS methods in Dataset~3 while varying the number of GPUs. Note that we tripled the amount of traces in the dataset to increase the use of GPUs in FO-CRS.

\begin{figure}[thb!]
\begin{center}
\includegraphics[width=1.00\columnwidth]{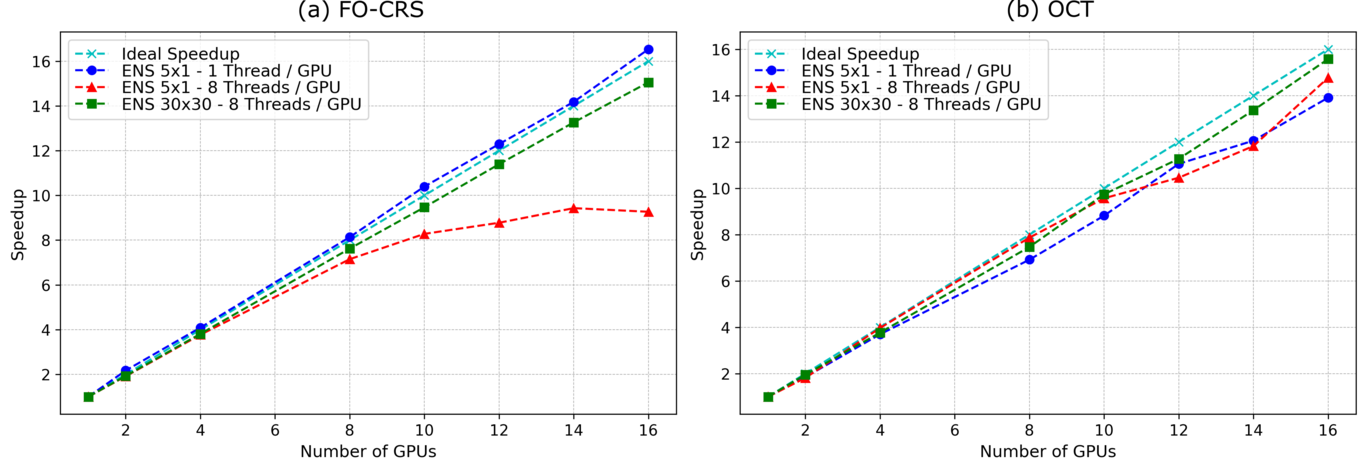}
\caption{{Scalability of Dataset~3 regarding the increase of computing resources (GPUs).
{\label{603036}}%
}}
\end{center}
\end{figure}

We compute the speedup in Figure~\ref{603036} as the execution time with one GPU compared to how many devices are used in the current execution. In an ideal scenario, the performance should increase linearly with the number of devices. Notably, the OCT execution, presented in Figure~\ref{603036}~(b), has an ideal speedup, with a slight decrease in linearity when going from ten to twelve GPUs on the $(5\times 1)$ setting with eight threads; however, it still presents performance gains. Because of the more time-consuming kernel, such linearity loss does not happen in the $30\times 30$ setting. Note that when the execution is performed with a single thread, thus reducing GPU load and increasing execution time, the speedup is again around the ideal line.

The FO-CRS execution highlights such performance restrictions more significantly, presented in Figure~\ref{603036}~(a). Note that the FO-CRS already presents a bottleneck when going from eight to ten GPUs with the $(5\times 1)$ setting with eight threads, achieving a broad plateau when going from fourteen to sixteen GPUs, in which there is no performance increase. However, perfect scalability is observed again if the $(5\times 1)$ configuration is performed with a single thread. Note that some points scale slightly above the ideal line, caused by run-to-run execution time variations. Such visual information confirms that the framework overhead introduces the bottleneck. Also, even though the scaling is significantly better with one thread than with eight threads, we emphasize that the execution time is significantly longer.

Unlike the results presented in the previous section in Figure~\ref{728261} with a single GPU, the faster execution times are not expected to decrease energy consumption.
Because increasing the number of GPU devices in the execution leads to higher power draw proportional to the number of devices added. Therefore, we would observe the exact overall energy requirements in an ideal scalability scenario. Ultimately, due to lower than linear scalability, we increase energy usage. However, it is a tiny margin compared to the execution time gains obtained. For example, in the OCT execution using the $(5\times 1)$ setting with eight threads, we observe an execution almost fifteen times faster for eight percent higher energy usage.

\section{Critics and limitations}

Although we have shown some positive results in favor of the ENS technique, it is still essential to present some known limitations. In this paragraph, we discuss these limitations, such as the issue of ENS convergence in larger configurations and the bucket size restriction for ENS-GPU.

\subsection{Qualitative limitations}

Figure~\ref{fig:rms_coherence} shows that ENS with interpolation had some issues with the quality compared to JADE. For instance, in Figure~\ref{fig:rms_coherence}~(a), JADE started producing slightly better quantitative coherence results after the $(50\times 50)$ setting. It can also be observed in Figure~\ref{fig:rms_coherence}~(b), in which JADE starts getting better after the $(40\times 40)$. We described such a limitation in the performance-quality trade-off by an interpolation process section. In order to analyze it, we make the difference between the Semblance panels obtained from performing JADE and ENS on Dataset~1 with an $(80\times 80)$ dataset and set the negative part of that subtraction to zero. Then we notice that, regardless of the traveltime operator shown, most of the better qualitative gains brought by JADE are, in fact, at the Semblance edge values over dip events.

Furthermore, we remove the ENS algorithm's interpolation step (i.e., $N_{skip} = 0$), allowing more samples to undergo forward and backward-and-forward propagation processes. When, again, subtracting the results obtained from such execution against JADE, we also conclude that the edge values difference is caused, in fact, by the interpolation. Therefore, the interpolation step introduces a trade-off between performance and quality.

\subsection{Bucket size}

\begin{figure}[thb!]
\begin{center}
\includegraphics[width=0.70\columnwidth]{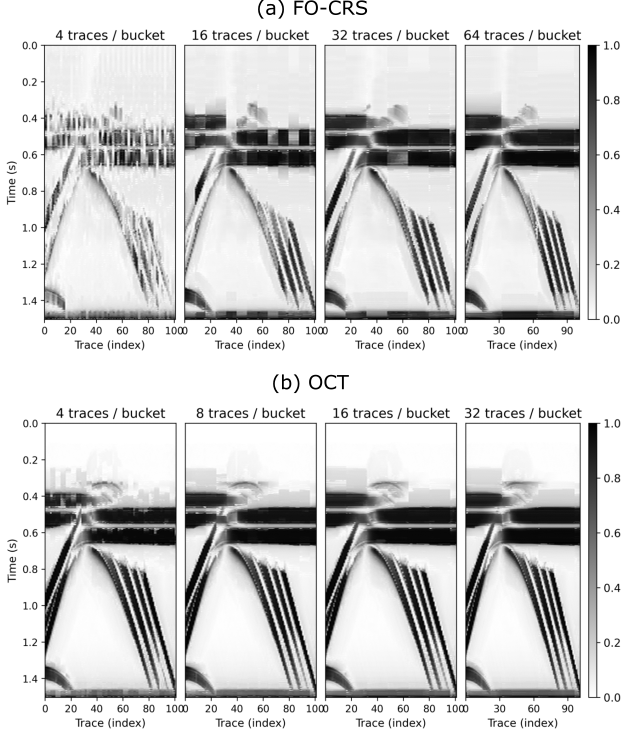}
\caption{{Dataset~1 - Comparison between (a) 4, 16, 32, and 64 traces per bucket in the FO-CRS $5\times 1$ estimation, and (b) 4, 8, 16, and 32 traces per bucket in the OCT $5\times 1$ estimation.
{\label{898636}}%
}}
\end{center}
\end{figure}

As already discussed, the number of domains in an interactive domain helps to reduce the number of searches without losing quality. Therefore, one of the most relevant parameters to the ENS execution is the bucket size since it can affect both scalability (for the ENS-GPU case) and qualitative results. To show this impact, let us consider Figure~\ref{898636}~(a), which presents the common-offset 500m from Dataset~1, obtained from executing an FO-CRS ENS estimation using five individuals and one generation but varying the bucket size. The execution with only four traces per bucket produced unacceptable results compared to a $(5\times 1)$ JADE execution. However, as the bucket size increases from 4 traces per bucket to 16, 32, and then 64, it is possible to see an improvement. This improvement happens because the parameters were allowed to be propagated more times in the execution with larger bucket sizes than those with smaller buckets. Therefore, the bigger bucket size allows the small number of individuals and generations to be compensated. Similarly, we have Figure~\ref{898636}~(b), which is related to an OCT estimation. However, unlike the previous results, the OCT executions required fewer traces inside their bucket to achieve acceptable results. Eight traces per bucket already seems to be satisfactory for that specific dataset. This convergence phenomenon is related to the OCT being more accurate than the FO-CRS.

Although bigger bucket sizes are expected to generate better results, it is important to highlight that the bucket size directly impacts the scalability. As buckets increase in size, the number of available buckets to be processed in parallel decreases. Therefore, restricting the benefits of using more GPUs and computing nodes in the execution.

Another essential detail is the total number of threads and GPUs per worker. After all, the received bucket by the worker is split through all the executing CPU threads. As each thread processes a subset of the bucket, the qualitative result is directly affected when we have too many threads and not large enough buckets. Therefore, it is necessary that for each hardware configuration, calculations are made so that the execution can produce the best trade-offs between scalability and satisfactory qualitative results.

\section{Discussion}

Besides the known limitations of the ENS technique, we must also consider data movement over our computations for the ENS-GPU approach. For this matter, we compare our technique's expected computational time over the real-time measured in our experiments. We then explore the execution time of each portion of our technique, aiming to understand the overall contribution of data movement between host and device.

\subsection{The number of Semblance computations}

By far, the most expensive operation in the whole evolutionary process is the Semblance computation. This happens because the complexity of all the other operations is constant time operation, while the Semblance call is linear with the number of traces in the specified fold of the traveltime, which is usually in the order of hundreds to thousands of traces, as shown by Table~\ref{tab:datasets}.

From a direct comparison between the Semblance calls made by each JADE and ENS (as shown by eqs.~(\ref{eq:jade_semblance}), (\ref{eqn:ens_semblance}), and the other ENS parameters previously setting), we should expect JADE-GPU to have an execution time at best equal to the ENS-GPU execution time since
\begin{equation}
\frac{num \underline \ ens \underline \ semblances \underline \ calls}{num \underline \ jade \underline \ semblances \underline \ calls} \le 1.0 \, , \text{as\ } NP_{f} \ge 5 \, , NG_{f} \ge 1\, .
\end{equation}
Also, for large population size and a number of generations, thus a considerable value of $NP_{f}\ \times\ \left(1\ +NG_{f}\right)$, we expect JADE-GPU to be almost three times slower. Besides, other configurations of $NP_{bf}$ and $NG_{bf}$ did not affect the quality results (to increase performance) on the studied dataset with $N_{skip}= 2$.

Figure~\ref{fig:ens_jade_execution_time} shows that for large configurations, the theory holds, especially in Dataset~3 with JADE-GPU being 2.76 and 2.66 times slower than ENS-GPU for FO-CRS and OCT, respectively. However, in smaller configurations, JADE-GPU is two times faster than ENS-GPU, therefore showing that there is a possible bottleneck and the performance is not defined by the number of semblances performed, which we discuss in the following subsection on data overhead.

\subsection{Data overhead}

Despite the Semblance calculation being an essential measure for the execution time, we must also consider the data movement overhead involved in the execution. First of all, the whole set of traces used in a Semblance computation must be migrated from the disk to the GPU memory before the evolutionary process of a given geometry trace begins. And, since the set of traces varies according to the current reference trace's offset and midpoint positions, there is a total of ``bucket size'' data movements on JADE-GPU and ``3 x bucket size'' data movements for ENS-GPU. Notably, the number of data transfers is directly proportional to the number of stages of each method, one for JADE-GPU and three for ENS-GPU, which makes ENS-GPU considerably more limited by the overhead introduced by such transfers than regular JADE-GPU. Also, another overhead that impacts the overall time is the number of kernel calls and how memory is accessed in the GPU (whether each GPU thread access contiguous data or not). Mainly, ENS-GPU is more sensitive to such overheads as well, thanks to performing more kernel calls due to the three stages and accessing data in a non-linear way, thanks to the neighboring approach. 

To measure the overhead impact on the execution, we used~\textbf{Nvprof} and~\textbf{Nsight System} 2021.1.3 platform of Nvidia to profile the code and obtain the overall time spent inside kernel calls and the overall time spent on memory calls (such as host to device movement and vice versa). Also, we considered only one GPU and only one computing CPU thread throughout the whole profiling. Then we present the obtained results in Tables~\ref{965355} and~\ref{543511}. The first table is associated with an OCT JADE-GPU parameter estimation of a slice of Dataset~3 (with only 1467 traces and 10MB). The second table is related to the OCT ENS-GPU estimation of the same dataset slice and traveltime. In both tables, we considered the whole execution time to produce the results for the $(5\times 1)$ and $(80\times 80)$ settings.

According to Tables~\ref{965355} and~\ref{543511}, it is evident that memory overhead is more impactful when settings are small ($5\times 1$). For instance, more than 30\% of the execution time is wasted outside the kernel calls. However, when the setting increases to $(80\times 80)$, the memory overhead becomes negligible, going to less than 4\% of the execution time. Interestingly, the time spent waiting for memory transfers does not increase alongside the increase in dataset size. This can be observed in Table~\ref{965355}, representing JADE-GPU, that the time spent in the memory column is around eight seconds regardless of setting size, and, similarly, in Table~\ref{543511}, representing ENS-GPU, that the wait for memory transfers is around twenty-three seconds.

Another consideration is that ENS-GPU's compute time is around four times more expensive than JADE-GPU in the $(5\times 1)$ setting, at around forty-seven seconds versus around twelve seconds. Such an event contradicts our previous calculations from Semblance computations. However, ENS-GPU has more overheads to be accounted for, namely the ones introduced by more kernel calls and access to non-contiguous memory, which explains the contradiction. In more significant scenarios, such as the $(80\times 80)$ set, the overhead becomes negligible, and JADE-GPU's computing time becomes around 1.77 times slower than ENS-GPU, at around one thousand versus around six hundred seconds.

\begin{table}[thb!]
\centering
\normalsize\begin{tabular}{|c|c|c|c|}
\hline
 & \textbf{Total Time}  & \textbf{Memory} & \textbf{Kernels}\\\hline
 \textbf{5x1} & 20 s (100 \%) & 8.2 s (41.0 \%) & 11.8 s (59.0 \%) \\\hline
 \textbf{80x80} & 1021 s (100 \%) & 7.15 s (0.7 \%) & 1013.8 s (99.3 \%) \\\hline
\end{tabular}
\caption{{JADE-GPU OCT Execution - Comparison between compute and overhead time for a slice of Dataset~3.
{\label{965355}}%
}}
\end{table}

\begin{table}[thb!]
\centering
\normalsize\begin{tabular}{|c|c|c|c|}
\hline
 & \textbf{Total Time}  & \textbf{Memory} & \textbf{Kernels} \\\hline
 \textbf{5x1} & 71 s (100 \%) & 23.6 s (33.2 \%) & 47.4 s (66.8 \%) \\\hline
 \textbf{80x80} & 597 s (100 \%) & 22.7 s (3.8 \%) & 574.3 s (96.2 \%) \\\hline
\end{tabular}
\caption{{ENS-GPU OCT Execution - Comparison between compute and overhead time for a slice of Dataset~3.
{\label{543511}}%
}}
\end{table}

Consider Tables~\ref{139581} and~\ref{645189} to delve deeper into this analysis. In Table~\ref{139581}, we depict the time taken to process an example trace of Dataset~3 using JADE-GPU and ENS-GPU with a setting of 5 individuals and 1 generation. Similarly, Table~\ref{645189} shows the execution time for that same trace but now considers an $(80\times 80)$ setting. Table~\ref{139581} shows that ENS-GPU has a slower execution time than JADE-GPU in smaller populations and generations settings. Note that ENS-GPU takes around the same time for each forward and backward estimation, with the forward estimation already being more expensive than JADE-GPU. Even though the last forward propagation is not shown in the table, its execution time is similar to the backward propagation; hence when adding up all kernel times and memory overheads, ENS-GPU takes about 3.9 times longer to compute the trace's parameters than JADE-GPU -- forty-three seconds against only eleven seconds.

\begin{table}[thb!]
\centering
\normalsize
\adjustbox{max width=\textwidth}{\begin{tabular}{l|cc|cc|cc|}
% \cline{2-7}
\hhline{~------}
 & \multicolumn{2}{c|}{\textbf{First Forward Propagation}} & \multicolumn{2}{c|}{\textbf{Backward Propagation}} & \multicolumn{2}{c|}{\textbf{Last Forward Propagation}} \\ \hhline{~------}
 & \multicolumn{1}{c|}{\textbf{kernels (ms)}} & \textbf{memory (ms)} & \multicolumn{1}{c|}{\textbf{kernels (ms)}} & \textbf{memory (ms)} & \multicolumn{1}{c|}{\textbf{kernels (ms)}} & \textbf{memory (ms)} \\ \hline
\multicolumn{1}{|c|}{\textbf{ENS 5x1}} & \multicolumn{1}{c|}{9} & 4.4 & \multicolumn{1}{c|}{7.6} & 7.5 & \multicolumn{1}{c|}{7.6} & 7.5 \\ \hline
\multicolumn{1}{|c|}{\textbf{JADE 5x1}} & \multicolumn{1}{c|}{6.8} & 4.5 & \multicolumn{1}{c|}{-} & - & \multicolumn{1}{c|}{-} & - \\ \hline
\end{tabular}}
\caption{{ENS-GPU's and JADE-GPU's kernel and memory transfer times for each step of the execution of a single trace using the $5 \times 1$ setting.
{\label{139581}}%
}}
\end{table}

Table~\ref{645189} shows how the execution time of the backward and last forward execution of ENS-GPU, plus the memory overhead, becomes negligible compared to the first forward execution of an $(80\times 80)$ setting. Also, reducing the number of time samples processed in the GPU to one-third $NT$ presents 1.7 times gains in the kernel execution time compared to JADE-GPU.

\begin{table}[thb!]
\centering
\normalsize
\adjustbox{max width=\textwidth}{\begin{tabular}{l|cc|cc|cc|}
\hhline{~------}
 & \multicolumn{2}{c|}{\textbf{First Forward Propagation}} & \multicolumn{2}{c|}{\textbf{Backward Propagation}} & \multicolumn{2}{c|}{\textbf{Last Forward Propagation}} \\ \hhline{~------}
 & \multicolumn{1}{c|}{\textbf{kernels (ms)}} & \textbf{memory (ms)} & \multicolumn{1}{c|}{\textbf{kernels (ms)}} & \textbf{memory (ms)} & \multicolumn{1}{c|}{\textbf{kernels (ms)}} & \textbf{memory (ms)} \\ \hline
\multicolumn{1}{|c|}{\textbf{ENS 80x80}} & \multicolumn{1}{c|}{318} & 5 & \multicolumn{1}{c|}{7} & 8 & \multicolumn{1}{c|}{7} & 8 \\ \hline
\multicolumn{1}{|c|}{\textbf{JADE 80x80}} & \multicolumn{1}{c|}{580} & 10 & \multicolumn{1}{c|}{-} & - & \multicolumn{1}{c|}{-} & - \\ \hline
\end{tabular}}
\caption{{ENS-GPU's and JADE-GPU's kernel and memory transfer times for each step
of the execution of a single trace using the 80 x 80 setting.
{\label{645189}}%
}}
\end{table}

Although those overheads considerably impact the execution time, ENS-GPU is still more attractive than JADE-GPU from a cost-benefit analysis. For instance, in Figure~{\ref{fig:rms_coherence}} (b), ENS-GPU applied to Dataset~3 with a $(5\times 1)$ setting produces the same RMS value as JADE-GPU using a $(30\times 30)$ setting, however, taking around four times less execution time. Furthermore, this cross-comparison between execution time is valid for all the other datasets and traveltimes. Another example is shown in Figure~\ref{fig:rms_coherence}~(a), in which the ENS-GPU algorithm applied to Dataset~3 produced in a $(5\times 3)$ setting the same results as JADE-GPU in a $(30\times 30)$ setting, with five times less execution time. Also, we must remember that ENS-GPU is cheaper to execute once it consumes much less energy, as previously shown. Therefore, despite comparing the execution time of small settings indicating that JADE-GPU is faster than ENS-GPU, it becomes clear that ENS-GPU is the best option to use in most scenarios when the comparison is brought to a fair instance.

\subsection{Mitigating the overhead}\label{mitigating-the-overhead}

As shown previously, memory overhead and execution time of the backward and forward propagation kernels can considerably decrease the gains brought by ENS-GPU, especially under small settings ($5\times 1$) that are of practical interest to us. To tackle the memory overhead, we already used two threads on the device so that we could partially overlap compute with memory operations. More specifically, at the beginning of every trace estimation, we launch two OpenMP threads, such that one is responsible for executing the ENS-GPU kernels. At the same time, the other is responsible for reading the seismic traces from the prestack data stored in the disk, which are used in the following trace estimation. The subsequent trace estimation has a decreased waiting time for its fold to be loaded from storage to system memory. Notably, we could also read the buffer directly from the disk to the GPU memory. However, that could significantly increase memory usage, as the fold of the subsequent trace estimation may not entirely overlap with the estimated current trace. Exploring this folding overlap still is an exciting topic for future works. However, the trade-off between memory transfers and memory usage would still be an issue. Nonetheless, we use asynchronous copy to transfer the data from CPU to GPU before the ENS-GPU kernels are launched, saving some memory transfer overhead time.

\section{Conclusion}

We have presented a brand new coevolution metaheuristic called ENS. This technique mainly uses the JADE algorithm to guide the estimation process toward convergence. However, it also leverages the data redundancy present in the searching dataset, ultimately speeding up the execution process. Qualitative results show that ENS achieves similar or superior outcomes to JADE but almost five times faster and with thirty-three less energy consumption. Also, the fact that a $(5\times 1)$ setting frequently produces acceptable results implies that we are almost on the verge of solving the traveltime parameter search problem. After all, the minimum amount of Semblance computations required per time sample is one, and our technique only computes ten Semblances per sample. Remarkably, ENS-GPU's most significant challenge involves the memory transfer overheads, which take about the same time as the kernel computation. For instance, ENS-GPU performs three kernel calls compared to a single one in JADE-GPU. Notably, each kernel call needs to move data between the system memory and the GPU device, with such transfer taking as long as the kernel computation for small settings. Therefore, plenty of room exists to improve the memory and compute overlap of the technique further. In conclusion, many opportunities exist to explore data redundancy in seismic parameter estimation. Nevertheless, the ENS technique presents itself as an interesting first approach, bringing significant improvements by offering multiple execution times and decreasing energy consumption over other search algorithms. As a first approach, we highlight that plenty of improvements could be made to the algorithm.

\backmatter

% \bmhead{Supplementary information}

% If your article has accompanying supplementary file/s please state so here. 

% Authors reporting data from electrophoretic gels and blots should supply the full unprocessed scans for key as part of their Supplementary information. This may be requested by the editorial team/s if it is missing.

% Please refer to Journal-level guidance for any specific requirements.

\bmhead{Acknowledgments}

The authors thank the High-Performance Geophysics (HPG) team for technical support.

\bibliography{sn-bibliography}% common bib file

\end{document}